\documentclass[reprint,aps,prd,nobibnotes,nofootinbib,showkeys,floatfix]{revtex4-2}
\usepackage{graphicx, caption, subcaption}
\usepackage[colorlinks=true, allcolors=blue]{hyperref}
\usepackage{etoolbox}
\AtBeginEnvironment{thebibliography}{\raggedright}
\usepackage{amsmath,amssymb,amsthm,amsfonts}
\usepackage{mathtools,mathrsfs,dsfont,bbold,slashed,xfrac}
\usepackage{multirow,array,makecell,booktabs,tablefootnote,dcolumn}
\usepackage{enumitem} 
\usepackage{soul} 
\usepackage{orcidlink}
\usepackage{tikz}
\usetikzlibrary{positioning,shapes.geometric,arrows.meta}

\begin{document}

\title{Search sensitivity for type-III seesaw heavy leptons in simulated \texorpdfstring{$pp$}{pp} collisions at \texorpdfstring{$\sqrt{s}=13.6$}{sqrt(s)=13.6} TeV at the LHC}

\author{A.M.M. Elsayed\orcidlink{0000000249554958}}
\affiliation{Department of Modern Physics and State Key Laboratory of Particle Detection and Electronics, University of Science and Technology of China, Hefei, China}

\author{Yusheng Wu\orcidlink{0000000215284865}}
\thanks{Corresponding authors: Yusheng Wu (\href{mailto:wuyusheng@ustc.edu.cn}{wuyusheng@ustc.edu.cn}) and A.M.M. Elsayed (\href{mailto:ammelsayed@mail.ustc.edu.cn}{ammelsayed@mail.ustc.edu.cn}).}
\affiliation{Department of Modern Physics and State Key Laboratory of Particle Detection and Electronics, University of Science and Technology of China, Hefei, China}

\date{\today}

\begin{abstract}
A search for the pair production of TeV-scale heavy leptons as predicted by the type-III seesaw mechanism is presented. The analysis is based on a simulated signal and Standard Model background data corresponding to an integrated luminosity of $500~\mathrm{fb}^{-1}$ of proton-proton collisions at $\sqrt{s}=13.6$ TeV. The ATLAS detector response is modeled with Delphes. Events are categorized based on the number of final state leptons and large-$R$ jets into nine mutually exclusive search channels. Multivariate analysis techniques are employed to separate the signal from the Standard Model backgrounds. Under the flavor-democratic mixing scenario ($\mathcal{B}_e=\mathcal{B}_\mu=\mathcal{B}_\tau=1/3$), the expected 95\% confidence level upper exclusion limit on the heavy lepton mass is approximately $1.6$ TeV.
\end{abstract}


\maketitle

\section{Introduction}\label{introduction}

The experimental observation of neutrino flavor oscillations provides conclusive evidence that neutrinos have nonzero masses that are much smaller than those of the charged leptons \cite{review_of_particle_physics}. While in the Standard Model (SM) the other fermions acquire masses by coupling to the Higgs boson, the origin of the neutrino masses remains uncertain. This is because the extremely small values of the neutrino masses compared to the masses of the other fermions appear unnatural in the SM. Several theoretical frameworks have been proposed to explain the smallness of neutrino masses, with the seesaw mechanism being among the most elegant and well-motivated. In its generic form the mechanism is realized in three variants, distinguished by the nature of the heavy fields introduced into the SM, namely a fermionic singlet in the type~I seesaw \cite{minkowski1977, gellmann1979, yanagida1979, mohapatra1980}, a scalar triplet in the type~II seesaw \cite{magg1980, schechter1980, lazarides1981, mohapatra1981}, and a fermionic triplet in the type~III seesaw \cite{foot1989, franceschini2008}. The smallness of the observed neutrino masses in the type~I and III seesaw mechanisms relies on the new massive states that couple to SM leptons and the Higgs field, whereas in the type-II seesaw the SM neutrino is assumed to couple solely to the new scalar triplet with a different vacuum expectation value. From a collider perspective, the type-II and III seesaw mechanisms are particularly attractive, as the new triplet fields can participate in electroweak gauge interactions. In this paper, a sensitivity study for the search for the heavy leptons predicted by the type-III seesaw mechanism is performed.

At the LHC, the type-III seesaw $\mathrm{SU}(2)_L$ fermionic triplets are pair-produced via electroweak (EW) interactions in both charged-charged ($N^+ N^-$) and charged-neutral ($N^\pm N^0$) modes. The heavy leptons undergo two-body decays into a SM boson ($W^\pm$, $Z$ or $H$) and a SM lepton ($\ell^\pm$, $\ell=e,\mu,\tau$). An example of a complete decay chain is $N^+ N^0 \to W^+\nu W^-\ell^+ \to \ell^+\ell^+\nu\nu jj$. The branching fraction ($\mathcal{B}_\ell$) for a heavy lepton decay into a specific light lepton flavor $\ell$ is proportional to $\big|V_\ell\big|^2 / \sum_\ell \big|V_\ell\big|^2$, where $V_\ell$ is the corresponding mixing matrix element. Electroweak precision measurements constrain these mixing angles to be $\lesssim 10^{-4}$ across all lepton flavors \cite{delaguila2008, biggio2020, das2021}. This upper bound is large enough to ensure prompt decays of heavy leptons in the mass ranges accessible at the LHC. Heavy state transitions are only relevant for $V_\ell < \mathcal{O}\left(10^{-8}\right)$ \cite{franceschini2008}, and consequently have a negligible effect in our search and are ignored. Throughout, the triplet components are assumed to be mass-degenerate, and their decays are prompt. The free parameters of the model are therefore the heavy lepton mass $m_N$ and the leptonic branching fractions $\mathcal{B}_e$, $\mathcal{B}_\mu$, and $\mathcal{B}_\tau$. In this analysis, we adopt the flavor-democratic mixing (FDM) scenario, $\mathcal{B}_e=\mathcal{B}_\mu=\mathcal{B}_\tau=1/3$, so that $m_N$ is the only remaining free parameter.

Previous searches at the LHC predominantly focused on multilepton final states for their low SM background contamination \cite{cms2012, atlas2015_8, atlas2015_9, cms2017, cms2020, atlas2021, atlas2022, cms2022}. The most stringent limits on the type-III seesaw model come from a 2022 CMS search \cite{cms2022} using the full LHC Run~2 dataset ($\sqrt{s} = 13$ TeV) focusing on multilepton final states with up to four light leptons and three hadronically decaying $\tau$ leptons. This search excluded type-III seesaw fermionic triplets with masses below $980$ GeV at 95\% CL under the FDM scenario, and below $990$, $1065$, and $890$ GeV in the exclusive electron, muon, and tau coupling scenarios, respectively. For heavier triplet fermions, the sensitivity of multilepton searches is limited by the rapidly decreasing pair-production cross section and by the small branching fractions of the leptonically decaying SM bosons. To extend the sensitivity to higher heavy-lepton masses, it is therefore increasingly important to exploit hadronic decays of the SM bosons in addition to the purely leptonic final states. 

\begin{figure}
\centering
\begin{subfigure}{0.4\textwidth}
\includegraphics[width=\textwidth]{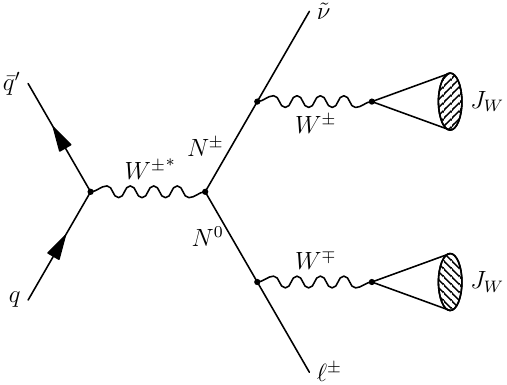}
\caption{}
\label{charged_neutral_production}
\end{subfigure}
\hfill
\begin{subfigure}{0.4\textwidth}
\includegraphics[width=\textwidth]{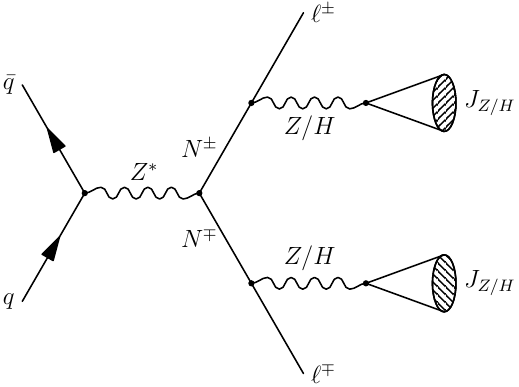}
\caption{}
\label{charged_charged_production}
\end{subfigure}
\caption{Example processes illustrating production and decay of type-III seesaw heavy lepton pairs at the LHC that result in final states containing multiple large-$R$ jets and leptons.}
\label{heavy_leptons_production}
\end{figure}

For TeV-scale heavy leptons, the decay products are typically highly boosted. The large Lorentz boost causes the hadronic decay products of the SM bosons to become highly collimated, such that they are preferentially reconstructed as a single large-$R$ jet rather than two resolved jets. Feynman diagrams illustrating example processes with final states containing leptons and large-$R$ jets are shown in Figure~\ref{heavy_leptons_production}. Phenomenological studies have explored the production and collider signatures of type-III seesaw heavy leptons, including multilepton channels and their prospects at the LHC \cite{franceschini2008, delaguila2009, biggio2012}. More recent work has considered multilepton final states with boosted hadronic boson decays at $\sqrt{s}=13$ TeV \cite{ashanujjaman2022}, highlighting the role of such final states at high $m_N$. In this work, we perform a systematic Monte Carlo study of the sensitivity to heavy leptons in simulated $pp$ collisions at $\sqrt{s}=13.6$~TeV, using $500~\mathrm{fb}^{-1}$ as an approximate integrated-luminosity benchmark for the combined LHC Run~2 and Run~3 data. In contrast to the previous phenomenological studies, we aim to improve the expected limits by exploring new search categories and employing advanced multivariate analysis techniques.

We define nine mutually exclusive channels spanning one-lepton, opposite-sign dilepton, same-sign dilepton, and multilepton final states, with and without reconstructed large-$R$ jets. These channels are chosen to capture most of the rare signal events and to allow the sensitivity to benefit from both leptonic decays and boosted hadronic decays of the EW bosons. For each channel, mass-parameterized boosted decision trees (pBDTs) \cite{pbdt} are used with a broad set of object-level and event-level features to discriminate the signal and background events. The resulting pBDT score distributions from all channels are then combined in a simultaneous likelihood fit to derive expected exclusion limits on the heavy-lepton mass $m_N$.

The remainder of this paper is organized as follows. Section~\ref{the_model} introduces the type-III seesaw framework. The production and decays of the heavy leptons at the LHC are discussed in Sections~\ref{production_modes} and \ref{decay_modes}, respectively, while the collider signatures and event categorization is discussed in Section~\ref{collider_signatures}. Section~\ref{monte_carlo_simulation} describes the Monte Carlo event generation, object reconstruction, and selection of leptons and large-$R$ jets. The event selection and analysis strategy are presented in Section~\ref{analysis_strategy}, while the expected sensitivity of this search is discussed in Section~\ref{results}. Finally, Section~\ref{conclusion} summarizes the results and presents the conclusions.

\section{Type-III Seesaw Model}\label{the_model}

The type~III seesaw model extends the SM matter content with three generations of leptonic $\mathrm{SU}(2)_L$ triplets that transform as $(\mathbf{1},\mathbf{3},0)$ under the SM gauge group $\mathrm{SU}(3)_c \times \mathrm{SU}(2)_L \times \mathrm{U}(1)_Y$. At least two triplet generations are required to reproduce the two observed neutrino mass-squared differences. Unless the triplet generations are nearly mass degenerate, the lightest triplet is expected to dominate the production rate and would therefore be the first to be discovered at the LHC \cite{biggio2012}. Therefore, we consider only one single generation of the heavy fermionic triplets for the purpose of this analysis. The triplet field is represented in the adjoint representation of $\mathrm{SU}(2)$ as follows:
\begin{equation}\label{sigma_triplets}
\Sigma \equiv \frac{1}{\sqrt{2}}\sigma_i\Sigma^i
=\begin{pmatrix}
\frac{1}{\sqrt{2}}\Sigma^0  &  \Sigma^+\\
\Sigma^-    &   -\frac{1}{\sqrt{2}}\Sigma^0
\end{pmatrix},
\end{equation}
where $\Sigma^\pm \equiv \frac{1}{\sqrt{2}}\big(\Sigma^1 \mp i\Sigma^2\big)$ and $\Sigma^0 \equiv \Sigma^3$ are two-component spinors with $\mathrm{U}(1)_\mathrm{EM}$ charges $Q=\pm1$ and $Q=0$, respectively. The conjugate triplet field can be written as
\begin{equation}\label{conjugate_sigma_triplets}
\Sigma^c \equiv \mathcal{C}\overline{\Sigma}^\mathsf{T}
=\begin{pmatrix}
\frac{1}{\sqrt{2}}\Sigma^{0c}  &  \Sigma^{-c}\\
\Sigma^{+c}    &   -\frac{1}{\sqrt{2}}\Sigma^{0c}
\end{pmatrix},
\end{equation}
where $\mathcal{C}$ is the charge-conjugation matrix. The Lagrangian involving the lepton triplet fields, along with the SM part denoted by $\mathscr{L}_\text{SM}$, is given by $\mathscr{L}=\mathscr{L}_\text{SM}+\mathscr{L}_\Sigma$, with \cite{biggio2012, abada2008}
\begin{multline}\label{triplets_lagrangian}
\mathscr{L}_\Sigma=
i \mathrm{Tr}\Big( \overline{\Sigma} \gamma^\mu D_\mu \Sigma \Big)
- \frac{1}{2} \mathrm{Tr} \Big(\overline{\Sigma} M_{\Sigma} \Sigma^c + \overline{\Sigma^c} M_{\Sigma}^* \Sigma\Big) \\
-\sqrt{2} \sum_{\alpha=e,\mu,\tau} \Big(\widetilde{H}^\dagger \overline{\Sigma} Y_{\Sigma\alpha} L_\alpha + \overline{L}_\alpha Y_{\Sigma\alpha}^\dagger \Sigma  \widetilde{H}  \Big),
\end{multline}
where $L_\alpha=\big(\nu_{\alpha{L}}, \ell_{\alpha{L}}\big)^\mathsf{T}$ are the SM left-handed lepton doublets, $H=\big(\phi^+,\phi^0\big)^\mathsf{T}$ is the SM Higgs doublet\footnote{$\widetilde{H}\equiv i\sigma_2H^*=\big(\phi^{0*},-\phi^-\big)^\mathsf{T}$}, $Y_{\Sigma}=\big(Y_{\Sigma e},  Y_{\Sigma\mu}, Y_{\Sigma\tau}\big)$ is the Yukawa coupling matrix, $M_\Sigma$ is the leptonic triplet mass matrix and $D_\mu$ is the gauge-covariant derivative given by
\begin{equation}
D_\mu = \partial_\mu -ig
\begin{pmatrix}
W_\mu^3 & \sqrt{2} W_\mu^+ \\
\sqrt{2} W_\mu^- & -W_\mu^3 \\
\end{pmatrix}.
\end{equation}
If the three triplet generations are assumed to be degenerate, then $M_\Sigma$ is proportional to $I_{3\times3}$. For a single triplet generation, $M_\Sigma$ reduces to a real mass parameter $m_\Sigma$. 
The charged triplet components ($\Sigma^\pm$) are combined into the Dirac spinor $\Psi = \Sigma^-_R + \Sigma^{+c}_R$ (with $\Psi_R = \Sigma_R^-$ and $\Psi_L = (\Sigma^{+}_R)^c$) to conveniently express the mixing of the SM charged leptons with the new triplet fields. The neutral component remains in the two-component notation. Throughout this work we adopt the convention of Refs. \cite{biggio2012,abada2008}, in which the spinors are chosen to be right-handed under Lorentz transformations. In this notation, the Lagrangian decomposes into $\mathscr{L}_\mathrm{kin} + \mathscr{L}_\mathrm{gauge}+ \mathscr{L}_Y$, where the first term, given by
\begin{multline}\label{kinetic_term}
\mathscr{L}_\mathrm{kin} = 
i\overline{\Psi}\slashed{\partial}\Psi 
+ i\overline{\Sigma^0_{R}}\slashed{\partial}\Sigma^0_{R} \\
- m_\Sigma\overline{\Psi}\Psi 
- \left(\frac{1}{2}m_\Sigma\overline{\Sigma^0_{R}}\Sigma^{0c}_{R} + \mathrm{h.c.}\right),
\end{multline}
describes the kinetic and mass terms of the triplet components. Note that both the charged and neutral triplet components already have a gauge-invariant bare mass terms before electroweak symmetry breaking (EWSB). The second term describes the interactions with the electroweak gauge bosons and is given by
\begin{multline}
\mathscr{L}_\mathrm{gauge} = 
- g W_\mu^3 \overline{\Psi}\gamma^\mu\Psi \\
+ g \big( 
    W_\mu^+ \overline{\Sigma^0_{R}} \gamma^\mu P_R \Psi 
    + W_\mu^+ \overline{\Sigma^{0c}_{R}} \gamma^\mu P_L \Psi
    + \mathrm{h.c.} 
    \big),
\end{multline}
where $g$ is the $\mathrm{SU(2)}_L$ coupling. The chiral projectors are $P_{L} = \frac{1}{2}(1-\gamma^5)$ and $P_{R} = \frac{1}{2}(1+\gamma^5)$. Finally, the Yukawa term is given by
\begin{multline}\label{yukawa_term}
\mathscr{L}_Y = 
 - \phi^0 \overline{\Sigma^0_{R}} Y_{\Sigma\alpha} \nu_{\alpha L}
 - \sqrt{2} \phi^0 \overline{\Psi} Y_{\Sigma\alpha} \ell_{\alpha L} \\
 - \phi^+ \overline{\Sigma^0_{R}}  Y_{\Sigma\alpha}  \ell_{\alpha L}
 + \sqrt{2} \phi^+ \overline{\nu^c_{\alpha L}}  Y_{\Sigma\alpha}^\dagger \Psi
 +\mathrm{h.c.},
\end{multline}
describes the Yukawa interactions between the heavy leptons and SM leptons. After EWSB, the neutral component of the Higgs doublet acquires a vacuum expectation value $\langle\phi^0\rangle=v_0/\sqrt{2}$ with $v_0\simeq246$~GeV. The neutral lepton mass Lagrangian terms can be collected into the following symmetric matrix as follows
\begin{equation}
\mathscr{L} \ni -\frac{1}{2}
\begin{pmatrix}
\overline{\nu_{\alpha L}} & \overline{\Sigma^{0c}_{R}}
\end{pmatrix}
\begin{pmatrix}
0 & m_{D\alpha}^{\mathsf T} \\
 m_{D\alpha} & m_\Sigma
\end{pmatrix}
\begin{pmatrix}
\nu^c_{\alpha L} \\
\Sigma^0_R
\end{pmatrix}
+\mathrm{h.c.}
\end{equation}
where $m_{D\alpha} = \frac{1}{\sqrt{2}} Y_{\Sigma\alpha} v_0$ is the Dirac mass term for the SM neutrinos. Similarly, the charged lepton mass terms can be written as
\begin{equation}
\mathscr{L}
\ni
-
\begin{pmatrix}
\overline{\ell_{\alpha L}} &
\overline{\Psi_{L}}
\end{pmatrix}
\begin{pmatrix}
m_\ell &
Y_{\Sigma\alpha}^\dagger v_0\\
0 &
m_\Sigma
\end{pmatrix}
\begin{pmatrix}
\ell_{\alpha R}\\
\Psi_R
\end{pmatrix}
+\mathrm{h.c.},
\end{equation}
where $m_\ell$ denotes the charged lepton mass matrix arising from the SM Higgs mechanism. The neutral mass matrix is symmetric because of the Majorana nature of the neutral fermions and is therefore diagonalized by a single unitary transformation. In contrast, the charged lepton mass matrix is a general Dirac mass matrix and is diagonalized by independent unitary transformations acting on the left- and right-handed fields. In the seesaw limit $m_{D\alpha} \ll m_\Sigma$, block diagonalization of the neutral mass matrix yields the well-known seesaw formula:
\begin{equation}\label{seesaw_formula}
m_\nu \simeq
-m_D^\mathsf{T}M_\Sigma^{-1}m_D
=
-\frac{v_0^2}{2}
Y_\Sigma^\mathsf{T}
M_\Sigma^{-1}
Y_\Sigma,
\end{equation}
where the smallness of the SM neutrino masses follows naturally from the large triplet mass scale $m_\Sigma$.  The diagonalization of the neutral and charged mass matrices also defines the physical heavy lepton mass eigenstates, denoted here by $N^0$ for the heavy Majorana neutrino and $N^\pm$ for the heavy charged leptons, both with mass $m_N$. Owing to their mixing with the SM leptons, these heavy states acquire interactions with the SM gauge and Higgs bosons. The production mechanisms and decay channels of these heavy states will be discussed in the following sections. Finally, it is convenient to express the Yukawa couplings in terms of the dimensionless mixing parameters:
\begin{equation}\label{mixing_paramaters}
V_\alpha = \frac{v_0}{\sqrt{2}} m_\Sigma^{-1} Y_{\Sigma\alpha}.
\end{equation}
The mixing parameters govern mixing-induced single production and the decays of the heavy leptons, while electroweak pair production is set primarily by the triplet gauge interactions and its mass and remains present when the mixing tends to zero \cite{biggio2012, franceschini2008}. They are constrained by electroweak precision measurements \cite{delaguila2008} and searches for flavor changing rare decays such as $\mu\to e\gamma$, $\tau\to\mu\gamma$ and $\tau\to e\gamma$ \cite{abada2008}.

\subsection{Production Modes}\label{production_modes}

\begin{figure}
\centering
\includegraphics[width=1\linewidth]{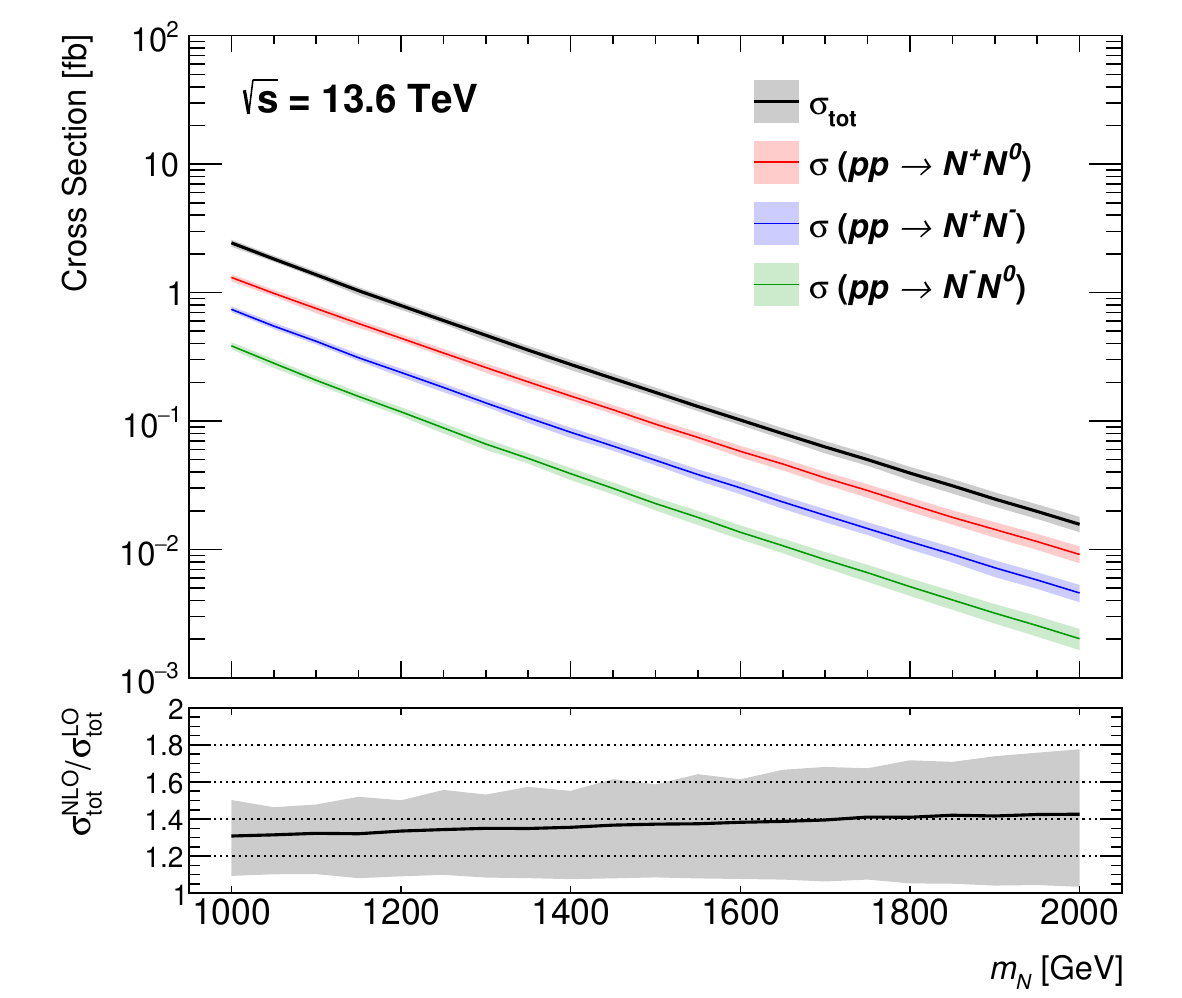}
\caption{Cross section for $pp \to N^+N^0$ (red), $pp \to N^+N^-$ (blue) and $pp \to N^-N^0$ (green) processes, varied with $m_N$, evaluated at NLO in QCD. NLO+PS and LO calculations are made with \textsc{MadGraph5\_aMC@NLO v3.7.2} \cite{madgraph1, madgraph2} with the \textsc{NNPDF4.0nlo} PDF sets using the \textsc{FeynRules} model of Ref.~\cite{typeiiiseesawNLO}. The uncertainty band includes variations in the renormalization and factorization scale, $\alpha_s$ and PDFs.}
\label{pair_production_cross_section}
\end{figure}

At the LHC, the type-III seesaw heavy leptons are produced predominantly through Drell-Yan processes\footnote{It is noteworthy that heavy lepton pairs are also produced, at much smaller rates, through EW vector-boson fusion (VBF) and gluon-gluon fusion processes. Direct production of neutral pairs $N^0N^0$ via $s$-channel $Z^*$ exchange is forbidden because $N^0$ has weak isospin $T_3=0$ and hypercharge $Y=0$. Thus, such a production mode exists only at strongly suppressed rates through $t$-channel $N^\pm$ exchange in $W^+W^-$ scattering.}. Charged pairs $N^+N^-$ are produced via the $s$-channel exchange of a virtual $Z/\gamma$, while the associated production $N^\pm N^0$ proceeds through $s$-channel $W^{*\pm}$ exchange as shown in Figures \ref{charged_charged_production} and \ref{charged_neutral_production}, respectively. Figure \ref{pair_production_cross_section} shows the cross section and its uncertainty for each production mode at next-to-leading (NLO) accuracy evaluated using \textsc{MadGraph5\_aMC@NLO} \cite{madgraph1, madgraph2} \textsc{v3.7.2} with the five-flavor scheme (5FS) and the \textsc{NNPDF4.0nlo} set of parton distribution functions (PDFs) \cite{nnpdf40, lhapdf}. The implemented \textsc{FeynRules} model is given in Ref. \cite{typeiiiseesawNLO}. The calculated cross-sections are compatible within uncertainties with the next-to-next-to-leading-order (NNLO) plus next-to-next-to-leading-logarithm (NNLL) accuracy calculations using $\mathrm{SU}(2)_L$ triplet production in an electroweak chargino-neutralino model \cite{resummino, susy_xsec}. As shown in Figure~\ref{pair_production_cross_section}, the cross sections fall rapidly with increasing $m_N$, reflecting the reduced phase space and the smaller parton densities probed at higher masses. The total production cross sections (NLO) for $m_N = 1.0$, $1.5$ and $2.0$ TeV are $2.438^{+0.167}_{-0.143}$ fb, $\left(16.681^{+1.324}_{-1.527}\right)\times 10^{-2}$ fb and $\left(1.568^{+0.209}_{-0.226}\right)\times 10^{-2}$ fb respectively.

\subsection{Decay Modes}\label{decay_modes}

\begin{figure}[tbp]
    \centering
    \includegraphics[width=\columnwidth,height=0.42\textheight,keepaspectratio]{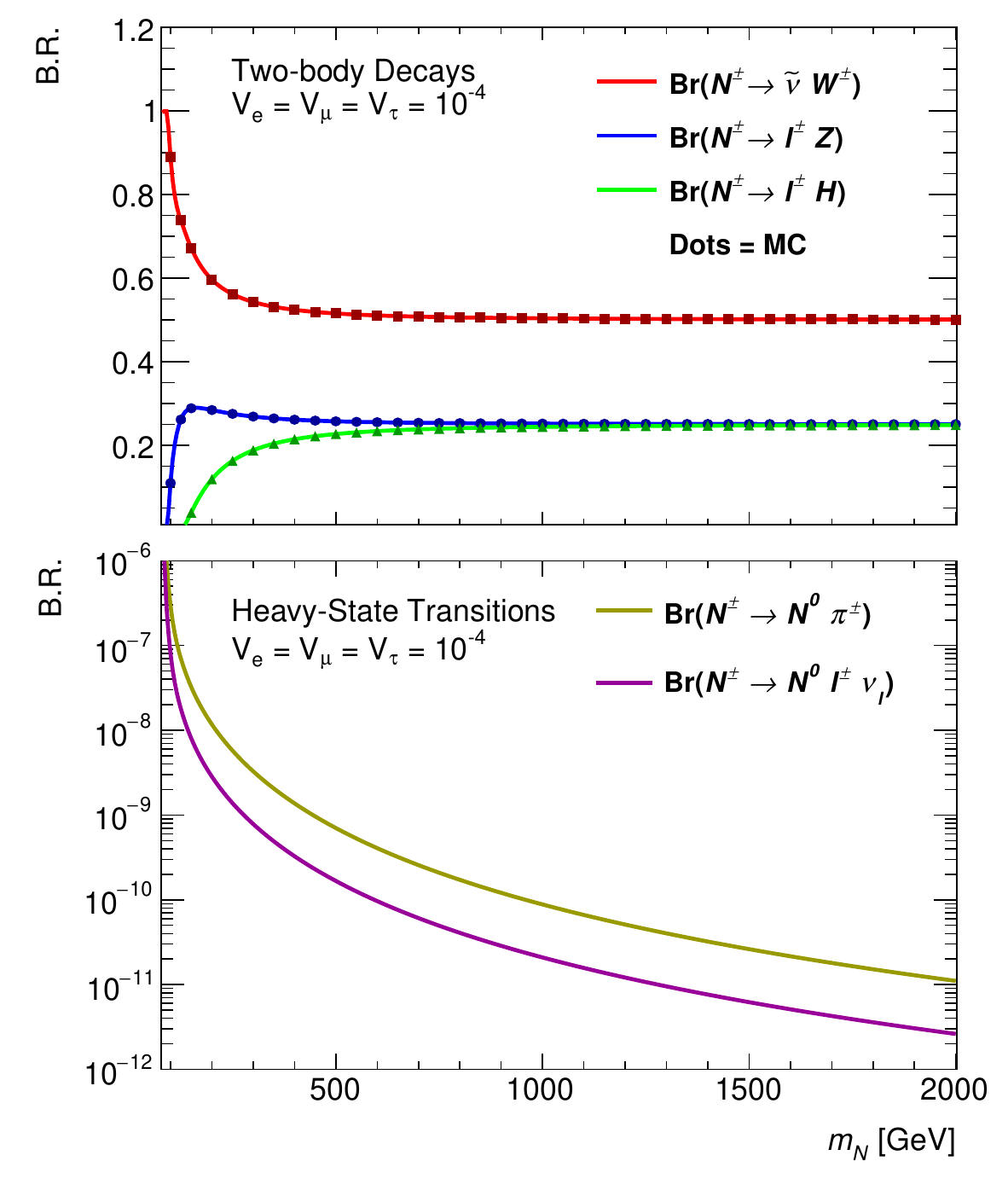}
    \caption{The branching ratio of heavy triplet fermion at different mass hypotheses under the flavor democratic mixing scenario (FDM) with $V_e=V_\mu=V_\tau = 10^{-4}$. The dots correspond to numerically evaluated values using \texttt{MadWidth} \cite{madwidth}, while the lines correspond to theoretical predictions \cite{biggio2012}. In the FDM scenario, the result is the same for the charged and neutral triplet decays.}
    \label{branching_ratios}
\end{figure}

The primary decay modes of $N^0$ and $N^\pm$ are two-body decays into a SM boson ($W^\pm$, $Z$ or $H$) and a SM lepton ($\ell_\alpha^\pm$ or $\nu_\alpha$). The allowed two-body decay modes are $N^0\to W^\pm \ell^\mp_\alpha$, $N^0\to Z \nu_\alpha$, $N^0\to H \nu_\alpha$ for the neutral component and $N^\pm \to W^\pm \nu_\alpha$, $N^\pm \to Z \ell^\pm_\alpha$, $N^\pm \to H \ell^\pm_\alpha$ for the charged components. The corresponding partial decay widths are given by \cite{franceschini2008, biggio2012}:
\begin{gather}
\Gamma\left(N^0 \rightarrow \ell_\alpha^{\pm} W^{\mp}\right) = \left|V_\alpha\right|^2 f_1\left(m_N, m_W\right), \label{N0_W_decays} \\
\sum_{\ell}\Gamma\left(N^0 \rightarrow \nu_\ell Z\right) = c_W^{-2} \sum_{\alpha}\left|V_\alpha\right|^2 f_1\left(m_N, m_Z\right), \label{N0_z_decays} \\
\sum_{\ell}\Gamma\left(N^0 \rightarrow \nu_\ell H\right) = \sum_{\alpha}\left|V_\alpha\right|^2 f_2\left(m_N\right), \label{N0_H_decays} \\
\sum_{\ell}\Gamma\left(N^{\pm} \rightarrow \nu_\ell W^{\pm}\right) = 2\sum_\alpha\left|V_\alpha\right|^2 f_1\left(m_N, m_W\right), \label{Npm_W_decays} \\
\Gamma\left(N^{\pm} \rightarrow \ell_\alpha^{\pm} Z\right) =c_W^{-2}\left|V_\alpha\right|^2 f_1\left(m_N, m_Z\right), \label{Npm_z_decays} \\
\Gamma\left(N^{\pm} \rightarrow \ell_\alpha^{\pm} H\right) = \left|V_\alpha\right|^2 f_2\left(m_N\right), \label{Npm_H_decays} 
\end{gather}
where
\begin{gather}
f_1\left(m_N, x\right) = \frac{g^2}{64 \pi}\frac{m_{N}^3}{x^2}\left(1-\frac{x^2}{m_{N}^2}\right)^2\left(1+2 \frac{x^2}{m_{N}^2}\right), \\
f_2\left(m_N\right) = \frac{g^2}{64 \pi}\frac{m_{N}^3}{m_W^2}\left(1-\frac{m_H^2}{m_{N}^2}\right)^2.
\end{gather}
Figure~\ref{branching_ratios} (top panel) displays the branching ratios of the two-body decays modes of the heavy leptons as a function of their mass in the flavor-democratic mixing scenario. The dots represent numerical evaluations obtained using \textsc{MadGraph5\_aMC@NLO} \cite{madgraph1, madgraph2} (specifically, \textsc{MadWidth} \cite{madwidth}) with the \textsc{FeynRules} model implementation of Ref.~\cite{typeiiiseesawNLO}, while the lines correspond to the theoretical predictions from equations \eqref{N0_W_decays} to \eqref{Npm_H_decays}. As expected, the Monte Carlo results are in perfect agreement with the analytical formulas.  The pattern of the branching ratios in Figure~\ref{branching_ratios} is dictated by the $\mathrm{SU}(2)_L$ symmetry and the Goldstone-equivalence theorem. At high masses, the phase space available to the $W^\pm$, $Z$ and $H$ channels becomes effectively identical, and the heavy states decay predominantly into the longitudinal polarizations of the electroweak bosons, which behave as the would-be Goldstone bosons. Since the Higgs doublet carries four real degrees of freedom; two for the longitudinal $W^\pm$, one for the longitudinal $Z$, and one for the physical Higgs, the decays become shared democratically, yielding an asymptotic ratio of $2:1:1$, that is $50\%$ to $W^\pm$, $25\%$ to $Z$ and $25\%$ to the $H$ decay channels. This pattern holds identically for both $N^0$ and $N^\pm$ because they reside in the same $\mathrm{SU}(2)_L$ triplet and their high-energy decays are governed uniformly by these scalar degrees of freedom. 

For a specific decay channel, a heavy lepton can decay into a lepton of flavor $\alpha$ with the branching fraction
\begin{equation}
\mathcal{B}_\alpha \propto \frac{\left|V_\alpha\right|^2}{\left|V_e\right|^2+\left|V_\mu\right|^2+\left|V_\tau\right|^2},
\end{equation}
set by the corresponding mixing element $V_\alpha$. This allows the collider phenomenology of the model to be fully specified by the heavy lepton mass $m_N$ together with the leptonic branching fractions $\mathcal{B}_e$, $\mathcal{B}_\mu$ and $\mathcal{B}_\tau$, without requiring precise knowledge of the overall Yukawa coupling scale. In the flavor-democratic mixing (FDM) scenario adopted in this study, the flavor branching fractions are fixed at $\mathcal{B}_e = \mathcal{B}_\mu = \mathcal{B}_\tau = 1/3$.

The components of the triplet are mass-degenerate at tree-level. One loop radiative corrections can lift the degeneracy. The dominant contributions come from loops involving virtual EW gauge bosons. Because the $N^\pm$ carries electric charge, the photon loop contributes exclusively to the self-energy of $N^\pm$, raising its mass slightly above $N^0$ by $\Delta m \approx 167$ MeV\footnote{Which also saturates at this value for $m_N > 500$ GeV} \cite{cirelli2006}. This splitting is large enough to open the heavy-state transition channels $N^\pm \to N^0 \pi^\pm$, $N^\pm \to N^0 e^\pm\nu_e$ and $N^\pm \to N^0 \mu^\pm\nu_\mu$. Because these decays are driven by the gauge interaction and the tiny mass splitting rather than the mixing, the emitted pion or lepton pair is very soft, and the collinear neutral state is produced at a measurable distance from the production point, giving a displaced secondary vertex\footnote{In fact, these kinds of decays result in a final states containing $N^0N^0$ from $pp\to N^\pm N^0$ when the pions are too soft and missed by the detector.}. The rate of these transitions scales as $\Delta m^3$ and is independent of the mixing parameters \cite{cirelli2006}, whereas the two-body decay rates scale as $\left|V_\alpha\right|^2$, so the relative importance of the two decay classes is controlled entirely by the size of the mixing parameters. As shown in the lower panel of Figure~\ref{branching_ratios}, at the benchmark of $|V_\alpha|=10^{-4}$ used in this analysis, which is consistent with the electroweak precision constraint $|V_\alpha|\lesssim 10^{-4}$ \cite{delaguila2008, biggio2020, das2021} and corresponds to the FDM scenario, the branching ratios of the heavy state transitions range from $\sim10^{-6}$ at $m_N = 100$ GeV down to $\sim10^{-12}$ at $m_N = 2.0$ TeV. Consequently, they are only relevant for mixing well below $\mathcal{O}\left(10^{-8}\right)$, where the two-body decay widths become sufficiently suppressed. At the benchmark mixing of $|V_\alpha|=10^{-4}$ adopted in this
analysis, the mixing-induced two-body decays dominate over the heavy-state transitions. These transitions become
relevant only for substantially smaller mixing parameters, a
regime that is not considered in the present analysis.

\subsection{Collider Signatures}\label{collider_signatures}

\begin{figure}
\centering
\includegraphics[width=\columnwidth,height=0.42\textheight,keepaspectratio]{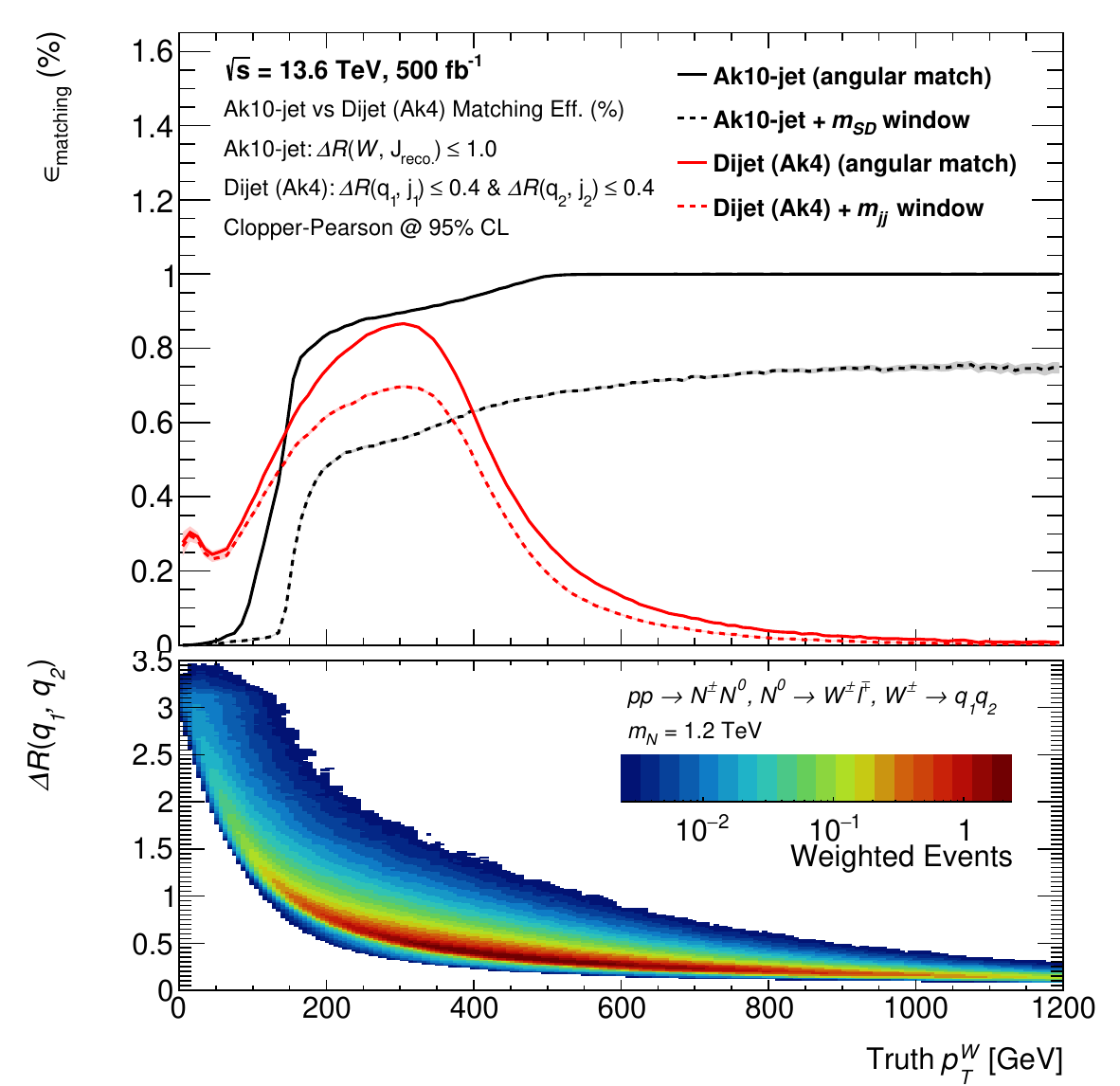}
\caption{Matching efficiencies for hadronically decaying $W$ bosons as a function of the truth boson transverse momentum. The lower panel shows the angular separation $\Delta R(q_1,q_2)$ of the two truth quarks, which follows $\Delta R(q_1,q_2) \simeq 2 m_W / p_{T}^{W}$. The upper panel shows the large $R$ matching efficiency in which the truth $W$ boson is matched to a large $R$ jet within $\Delta R \leq 1.0$, without (solid black line) and with an additional soft drop mass requirement $|m_\text{SD}-m_W| < 25$~GeV (dashed black line). This is compared with the resolved dijet reconstruction in which each truth parton is matched to a distinct small radius jet with $p_T > 25$~GeV within $\Delta R \leq 0.4$ (solid red line), without and with an additional dijet mass requirement $|m_{jj}-m_W| < 25$~GeV (bashed red line). Error bands denote the Clopper--Pearson intervals at 95\% confidence level.}
\label{fatjet_reconstruction_efficincy}
\end{figure}

For a heavy lepton of TeV scale mass the decay products are strongly boosted, and the two quarks from a hadronically decaying $W$, $Z$ or $H$ boson are emitted within an angular separation $\Delta R(q_1,q_2) \simeq 2m_{B}/p_T^{B}$, where $m_{B}$ and $p_T^{B}$ are the mass and transverse momentum of the decaying boson. Once $p_T^{B} > 2m_B$, the two quarks are more efficiently captured by a single large $R$ jet than by two resolved small $R$ jets. 
Throughout this paper large $R$ jets with radius parameter $R=1.0$ are called Ak10 jets\footnote{Similarly, ordinary jets (including QCD, $b,c,s$ and $\tau$ jets) clustered using the anti-$k_T$ algorithm with the radius paramater $R=0.4$ are hereafter referred to to as Ak4 jets.}. Figure~\ref{fatjet_reconstruction_efficincy} shows the matching efficiency of both strategies as a function of the true boson transverse momentum. The efficiency for matching a hadronically decaying $W$ boson to an Ak10 jet within $\Delta R \le 1.0$ rises steeply with $p_T^W$, exceeding $80\%$ above $200$~GeV and then reaches $100\%$ at $p_T^W \gtrsim 500$~GeV. Requiring in addition that the soft-dropped mass lie within $25$~GeV of $m_W$ lowers the efficiency, which then saturates at about $75\%$ at $p_T^W \gtrsim 600$~GeV. The resolved dijet efficiency, obtained by matching each true parton to a distinct small radius jet with $p_T>25$~GeV within $\Delta R\le0.4$, behaves in the opposite way. It peaks near $86\%$ at $p_T^W \simeq 300$~GeV and falls below $10\%$ above about $700$~GeV. Similarly, requiring $|m_{jj}-m_W| < 25$ GeV for the dijet mass $m_{jj}$ lowers the matching efficiency. For $m_N \ge 1.0$ TeV considered in this analysis, the SM bosons and leptons produced in the decays typically carry transverse momenta above $400$~GeV at matrix element level. In this regime the merged topology, in which the two quarks of a hadronic boson decay are captured by a single large $R$ jet, is the relevant one. Adding the resolved topology in the low-$pT$ regions, in which the two quarks are reconstructed as separate small radius jets, would increase the complexity of the analysis without a commensurate gain in sensitivity.

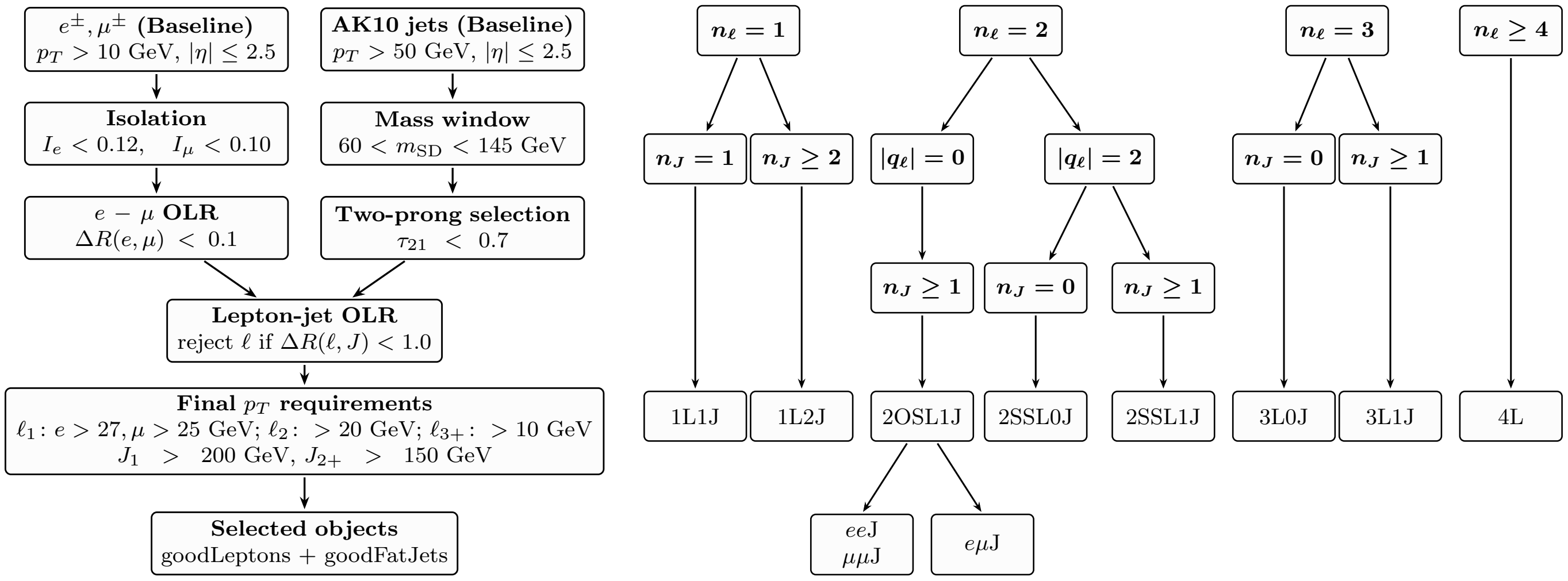
\begin{figure*}[tbp]
\centering
\begin{subfigure}[b]{0.385\textwidth}
\centering
\resizebox{\linewidth}{!}{%
\begin{tikzpicture}[
    font=\footnotesize,
    box/.style={rectangle, rounded corners=2pt, draw=black, line width=0.6pt, fill=black!1, align=center, text width=3.2cm, minimum height=0.8cm, inner sep=2.5pt},
    arrow/.style={-{Stealth[length=1.6mm]}, line width=0.7pt, shorten >=0.8pt, shorten <=0.8pt}
]
\node[box] (l1) at (-1.90,0)     {\textbf{$e^\pm,\mu^\pm$ (Baseline)}\\ $p_T>10$~GeV, $|\eta|\leq 2.5$};
\node[box] (l2) at (-1.90,-1.21) {\textbf{Isolation}\\ $I_e<0.12, \quad I_\mu<0.10$};
\node[box] (l3) at (-1.90,-2.42) {\textbf{$e-\mu$ OLR}\\ $\Delta R(e,\mu)<0.1$};

\node[box] (j1) at (1.90,0)      {\textbf{AK10 jets (Baseline)}\\ $p_T>50$~GeV, $|\eta|\leq 2.5$};
\node[box] (j2) at (1.90,-1.21)  {\textbf{Mass window}\\ $60<m_\mathrm{SD}<145$~GeV};
\node[box] (j3) at (1.90,-2.42)  {\textbf{Two-prong selection}\\ $\tau_{21}<0.7$};

\node[box, text width=3.35cm] (olr) at (0,-3.74) {\textbf{Lepton-jet OLR}\\ reject $\ell$ if $\Delta R(\ell,J)<1.0$};
\node[box, text width=7.5cm] (pt)  at (0,-5.03) {\textbf{Final $p_T$ requirements}\\ $\ell_1\colon e>27, \mu>25$~GeV; $\ell_2\colon >20$~GeV; $\ell_{3+}\colon >10$~GeV\\ $J_1>200$~GeV, $J_{2+}>150$~GeV};
\node[box, text width=3.75cm] (out) at (0,-6.47) {\textbf{Selected objects}\\ goodLeptons + goodFatJets};

\draw[arrow] (l1) -- (l2);
\draw[arrow] (l2) -- (l3);
\draw[arrow] (j1) -- (j2);
\draw[arrow] (j2) -- (j3);
\draw[arrow] (l3) -- (olr);
\draw[arrow] (j3) -- (olr);
\draw[arrow] (olr) -- (pt);
\draw[arrow] (pt) -- (out);

\end{tikzpicture}%
}
\caption{Object selection}
\label{fig:object_selection}
\end{subfigure}
\hfill
\begin{subfigure}[b]{0.59\textwidth}
\centering
\resizebox{\linewidth}{!}{%
\begin{tikzpicture}[
    font=\small,
    req/.style={rectangle, rounded corners=2pt, draw=black, line width=0.6pt, fill=black!1, align=center, text width=1.3cm, minimum height=0.7cm,  inner sep=2pt},
    chan/.style={rectangle, rounded corners=2pt, draw=black, line width=0.6pt, fill=black!1, align=center, text width=1.3cm, minimum height=0.7cm, inner sep=2pt},
    arrow/.style={-{Stealth[length=1.4mm]}, line width=0.7pt, shorten >=0.8pt, shorten <=0.8pt}
]
\node[req] (n1) at (0.85,0)   {$\boldsymbol{n_\ell=1}$};
\node[req] (n2) at (4.55,0)   {$\boldsymbol{n_\ell=2}$};
\node[req] (n3) at (9.15,0)   {$\boldsymbol{n_\ell=3}$};
\node[req] (n4) at (11.6,0)   {$\boldsymbol{n_\ell\geq4}$};

\node[req] (j11) at (0.10,-1.81)  {$\boldsymbol{n_J=1}$};
\node[req] (j12) at (1.60,-1.81)  {$\boldsymbol{n_J\geq2}$};

\node[req] (qo)  at (3.3,-1.81)   {$\boldsymbol{|q_\ell|=0}$};
\node[req, text width=1.4cm] (qs)  at (5.8,-1.81)   {$\boldsymbol{|q_{\ell}|=2}$};

\node[req] (j31) at (8.4,-1.81)   {$\boldsymbol{n_J=0}$};
\node[req] (j32) at (9.9,-1.81)   {$\boldsymbol{n_J\geq1}$};

\node[req] (jo1) at (3.3,-3.63)   {$\boldsymbol{n_J\geq1}$};
\node[req] (js0) at (4.9,-3.63)   {$\boldsymbol{n_J=0}$};
\node[req] (js1) at (6.7,-3.63)   {$\boldsymbol{n_J\geq1}$};

\node[chan] (c1) at (0.10,-5.44)  {1L1J};
\node[chan] (c2) at (1.60,-5.44)  {1L2J};
\node[chan] (c5) at (3.3,-5.44)   {2OSL1J};
\node[chan] (c6) at (4.9,-5.44)   {2SSL0J};
\node[chan] (c7) at (6.7,-5.44)   {2SSL1J};
\node[chan] (c3) at (8.4,-5.44)   {3L0J};
\node[chan] (c4) at (9.9,-5.44)   {3L1J};
\node[chan] (c8) at (11.6,-5.44)  {4L};

\node[chan, minimum height=0.85cm] (r5a) at (2.45,-7.25) {$ee$J\\$\mu\mu$J};
\node[chan, minimum height=0.85cm] (r5b) at (4.15,-7.25) {$e\mu$J};

\draw[arrow] (n1) -- (j11);
\draw[arrow] (n1) -- (j12);
\draw[arrow] (n2) -- (qo);
\draw[arrow] (n2) -- (qs);
\draw[arrow] (n3) -- (j31);
\draw[arrow] (n3) -- (j32);
\draw[arrow] (n4) -- (c8);

\draw[arrow] (qo) -- (jo1);
\draw[arrow] (qs) -- (js0);
\draw[arrow] (qs) -- (js1);

\draw[arrow] (j11) -- (c1);
\draw[arrow] (j12) -- (c2);
\draw[arrow] (jo1) -- (c5);
\draw[arrow] (js0) -- (c6);
\draw[arrow] (js1) -- (c7);
\draw[arrow] (j31) -- (c3);
\draw[arrow] (j32) -- (c4);

\draw[arrow] ([xshift=-0.20cm]c5.south) -- (r5a.north);
\draw[arrow] ([xshift= 0.20cm]c5.south) -- (r5b.north);

\end{tikzpicture}%
}
\caption{Event selection and channel definitions.}
\label{fig:channel_definition}
\end{subfigure}

\caption{Object selection and channel definition. (a) Sequential object selection applied to baseline electrons, muons, and AK10 jets. (b) Event classification tree showing the orthogonal search channels based on lepton multiplicity $n_\ell$, sum of leptons charge $q_\ell$, and Ak10 jet multiplicity $n_J$. The 2OSL1J channel is explicitly split into same-flavor and opposite-flavor regions to account for the different background compositions.}
\label{object_selection_and_channels}
\end{figure*}

Given that the dominant production mode is $pp\to N^+N^0$ (see Section~\ref{production_modes}), and that both heavy leptons decay mostly via the $W$ channel (see Section~\ref{decay_modes}), the most common final state contains one prompt lepton and two hadronically decaying $W$ bosons, $pp\to N^+N^0\to W^\pm_\mathrm{had}W^\pm_\mathrm{had}\ell^\mp\nu$ (see Figure~\ref{charged_neutral_production}), referred to as the 1L-2J channel, where ``J'' stands for Ak10 jet corresponding to a hadronically decaying SM boson. At parton level, this channel has most of the signal events. At reconstruction level, this channel becomes less populated by signal events compared to the 1L-1J channel due to lower Ak10 jet reconstruction and selection efficiencies compared to that of muons and electrons, causing a sizable fraction of signal events to lose one of the two Ak10 jets. The 1L-1J channel also gains contributions from channels with higher numbers of matrix-element level leptons when one or more leptons are missed by the detector, making it the one with the highest cross section\footnote{It is noteworthy that a fully hadronic ($0\ell$) topology, dominated by two hadronically decaying SM bosons in the final state, is also possible at parton level, and become dominant at reconstruction level. However, the overwhelming SM background makes it the most difficult channel to exploit and is therefore not considered in this analysis.}. 

The dilepton final states are next most probable, and split into opposite sign (OS) and same sign (SS) pairs. The OS dilepton channel, referred to as 2OSL, is the larger of the two, receiving contributions both from $N^+N^-$ production and from $N^\pm N^0$ events in which the $Z$ boson decays leptonically. The SS channel, referred to as 2SSL, is smaller but of particular theoretical interest, as it is a direct collider consequence of the Majorana mass term of Eq.~\eqref{kinetic_term}. The lepton number violating process $N^\pm N^0\to\ell^\pm Z\,\ell^\pm W^\mp\to\ell^\pm\ell^\pm J_Z J_W$ provides a particularly distinctive signature associated with the Majorana nature of the neutral heavy lepton, with comparatively small SM background contributions. Dominant background contributors for this channel are the $W^\pm W^\pm jj$, triboson and $t\bar{t}V$ processes. At the decay level, most dilepton events contain two hadronically decaying SM bosons. However, because one or both of their hadronic decays may fail the reconstruction or selection requirements, we require only one Ak10 jet, as shown in Figure~\ref{fig:channel_definition}. Similarly to the one-lepton (1L) channels case, a substantial fraction of the second Ak10 jets are not reconstructed or fail the selection requirements, therefore demanding a second jet rejects signal events without a commensurate gain in purity. The opposite sign dilepton channel is defined by requiring at least one Ak10 jet, refered to as the 2OSL-1J channel. To account for the different background composition, this channel is further split by the lepton flavors into $ee$J, $\mu\mu$J and $e\mu$J, where the $ee$J, $\mu\mu$J final states are merged and considered together (2OSSFL-1J channel). The same sign dilepton channel (2SSL) is split into a category with no Ak10 jets (2SSL-0J), which retains the events in which the hadronic decay products are missed or fall outside the acceptance, and one with at least one (2SSL-1J) as shown in Figure~\ref{fig:channel_definition}.

Final states with three or more leptons occur at lower rates, owing to the extra suppression from a second leptonic boson decay, but benefit from a very clean SM background and have historically been the primary target of LHC searches for this model \cite{cms2012, atlas2015_8, atlas2015_9, cms2017, cms2020, atlas2021, atlas2022, cms2022}. A representative trilepton final state is $N^\pm N^0\to\ell^\pm Z\,\ell^\pm W^\mp\to\ell^\pm\ell^\pm\ell^\mp\nu J_Z$. At parton level the model can produce final states with up to six leptons, and more if the $H\to4\ell$ decay is included, but the cross section falls steeply with the lepton multiplicity, and no previous LHC analysis has considered exclusive search channels with more than four leptons. In this analysis we define channels with exactly three leptons (3L channel), split into 3L-0J and 3L-1J by the requirements $n_J=0$ and $n_J\ge1$, and a channel with at least four leptons for which no requirement on the large $R$ jet multiplicity is imposed as shown in Figure~\ref{fig:channel_definition}. The analysis comprises nine search channels in total. 

\section{Monte Carlo Simulation}\label{monte_carlo_simulation}

\subsection{Event Generation}

This analysis is based on Monte Carlo simulation of signal and SM background events in $pp$ collisions at $\sqrt{s}=13.6$~TeV, using $500~\mathrm{fb}^{-1}$ as an approximate integrated-
luminosity benchmark representative of the combined LHC Run~2
and Run~3 datasets. For the signal events, we consider all 27 distinct leading order (LO) production and decay processes resulting from the decays of each of $N^+N^0$, $N^+N^-$ and $N^-N^0$ pairs into nine different pairs of $W^\pm$, $Z$ and $H$ bosons. Since the neutral component $N^0$ is a Majorana neutrino, both the $N^0\to \ell^+ W^-$ and $N^0\to \ell^- W^+$ channels are included with equal rates. Parton-level events are generated with \textsc{MadGraph5\_aMC@NLO} \cite{madgraph1, madgraph2} \textsc{v3.7.2} using the five-flavor scheme (5FS) and the \textsc{NNPDF4.0lo} set of parton distribution functions (PDFs) \cite{nnpdf40, lhapdf}. The implemented \textsc{FeynRules} model used for this analysis is given in Ref. \cite{typeiiiseesawNLO}. Different signal samples are generated for different mass hypotheses in the range $[1.0,2.0]$ TeV with a mass step of $50$ GeV.

We simulate and include all SM background processes giving rise to one or more prompt leptons. Such processes include $tW$, $t\bar{t}$, $t\bar{t}V$, diboson ($VV$), triboson ($VVV$) and $V+\mathrm{jets}$, where $V=W^\pm, Z$. To account for the same-sign dilepton channels, the $W^\pm W^\pm j j$ processes are also generated considering both QCD and EW contributions. The parton-level events are generated with \textsc{MadGraph5\_aMC@NLO} \cite{madgraph1, madgraph2} \textsc{v3.7.2} at LO using the 5FS and \textsc{NNPDF4.0lo} set of PDFs \cite{nnpdf40, lhapdf}. The top quarks and EW gauge bosons are decayed at matrix-element level. To focus the event generation in the high-$p_T$ phase space where the signal events live, the following generator-cuts were applied on the final state particles: $p_T(\ell_1) > 150$ GeV, $p_T(\ell_2) > 50$ GeV and $|\vec{p}_T^\mathrm{~miss}| > 50$ GeV, where $\ell_1$ and $\ell_2$ represent the leading and sub-leading leptons in $p_T$, respectively, and $|\vec{p}_T^\mathrm{~miss}|$ represents the missing transverse momentum. To avoid biasing the results due to these generator-level cuts, tighter preselection cuts are applied in all channels, as described in Section~\ref{analysis_strategy}.

The signal and background samples are interfaced to \textsc{Pythia 8.3} \cite{pythia} to model the parton shower and hadronization, with parameters set according to the \textsc{A14} tune \cite{a14_tune} and using the \textsc{NNPDF2.3lo} set of PDFs \cite{nnpdf23}. The $W^\pm+$jets and $Z/\gamma^*+$jets processes were generated in association with up to two jets followed by MLM matching in \textsc{Pythia}. The background samples are normalized to at least the NLO prediction \cite{th_tw_136, atlas_ttbar_136, th_ttw_136, th_ttz_136, th_ssww_136, cms_ww_136, cms_wz_136, atlas_zz_136, atlas_vjets_136, cms_vjets_136}.

Nonprompt or fake leptons and charge misidentification can arise from detector-reconstruction effects \cite{atlas_fake_leptons_2023, atlas_electron_id_2019}. Tight selection requirements can largely suppress contributions from nonprompt leptons without affecting the signal. The electron charge misidentification background is only relevant for the 2SSL search category. These can be largely suppressed by applying charge-quality requirements and a $Z$-veto to suppress the charge-misidentified $Z\to ee$ backgrounds \cite{cms_hh}. The probability of mismeasuring the charge of muons is negligible \cite{cms_tth}. Dedicated estimates of these effects are therefore outside the scope of the present phenomenological sensitivity study and are appropriate to a future analysis using collision data, where they can be constrained with data control samples.

\subsection{Object Reconstruction and Selection}

For all samples, a fast simulation of the \textsc{ATLAS} detector \cite{atlas_experiment_1, atlas_experiment_2} response was performed using \textsc{Delphes 3} \cite{delphes} configured for the LHC~Run~3 \cite{atlas_experiment_2}. The reconstructed electron and (muon) candidates are required to have transverse momentum $p_T>10$ GeV and must lie within the tracking system acceptance pseudo-rapidity $|\eta| < 2.5$ ($|\eta| < 2.7$). The relative lepton isolation ($I$) is required to be smaller than 12\% (15\%) for electrons (muons), in order to suppress backgrounds from jets misidentified as leptons. For each reconstructed electron or muon ($\ell$), the isolation variable ($I$) is defined as \cite{delphes}
\begin{equation}
I(\ell) = \frac{1}{p_T{(\ell)}} \sum_{i\neq\ell}^{\Delta R<R,~ p_T(i)>p_T^\mathrm{min}} p_T(i),
\end{equation}
where the numerator is the scalar sum of $p_T$ of all particles (with $p_T^\mathrm{min}= 0.5$ GeV) within a cone of $\Delta R = \sqrt{(\Delta\eta)^2 + (\Delta\phi)^2}=0.5$ around $\ell$, and the denominator is the $p_T$ of the lepton of interest.

Constituents of the large-$R$ jets are clustered using the anti-$k_T$ algorithm \cite{antikt}, with the radius parameter $R=1.0$, chosen to capture the full decay products of highly boosted massive particles while minimizing contamination from nearby soft radiation \cite{atlas_fatjet_7TeV, atlas_fatjet_13TeV}. Ak10 jet candidates must lie within the tracking system acceptance pseudorapidity $|\eta| < 2.5$, and are required to have transverse momentum $p_T>50$ GeV. For the signal events, Ak10 jets candidates are required to have $p_T>100$ GeV. To suppress soft and wide-angle QCD emissions, Ak10 jets are processed with the soft-drop grooming algorithm \cite{softdrop}. The soft-drop algorithm is applied with parameters $\beta=0$ and symmetry cut $z_{\rm cut}=0.1$, which systematically removes the soft wide-angle radiation, preserving the hard core of the jet while improving the discrimination of boosted objects. The algorithm parameters are chosen to optimize the selection following \cite{atlas_fatjet_13TeV}. To characterize the internal structure of the Ak10 jets, the $N$-subjettiness $\tau_{N}$ \cite{n_subjettiness1, n_subjettiness2} is used to probe their multi-prong nature. The one-pass winner-take-all $k_T$ axis is chosen for the minimization procedure, with thrust parameter $\beta = 1$. The ratio $\tau_{N}/\tau_{N-1}$ serves as a discriminant between $N$-prong and $(N-1)$-prong jets. In our case, the two-prong Ak10 jets from highly boosted SM bosons tend to have lower $\tau_{21} \equiv \tau_2 / \tau_1$ compared to one-prong Ak10 jets. 

Following object reconstruction, baseline electrons and muons are selected with $p_T>10$ GeV and $|\eta|< 2.5$. As shown in Figure~\ref{fig:object_selection}, the baseline lepton collection is refined using relative-isolation requirements of $I<12\%$ for electrons and $I<10\%$ for muons. To avoid double counting, electron-muon overlap removal is applied, with the lower-$p_T$ candidate discarded for pairs satisfying $\Delta R(e^\pm,\mu^\pm)<0.1$. Baseline Ak10 jets are selected with $p_T>50$ GeV and $|\eta|< 2.5$. Their soft-dropped mass is required to satisfy $m_W-m_\text{Win}<m_{\mathrm{SD}}<m_H+m_\text{Win}$ GeV, where $m_\text{Win}=20$ GeV, together with $\tau_{21}<0.7$, to enhance the contribution of two-prong jets consistent with boosted hadronic decays of electroweak bosons and suppress the predominantly one-prong QCD background. 

A further overlap-removal requirement is imposed between the selected leptons and Ak10 jets. Leptons within $\Delta R(\ell^\pm,J)<1.0$ of an Ak10 jet are rejected, thereby suppressing contamination from leptons associated with the hadronic activity contained in the large-$R$ jet. The remaining objects are required to satisfy transverse-momentum thresholds according to their ordering in $p_T$. The leading electron (muon) is required to satisfy $p_T>27$ ($25$) GeV, while the subleading lepton must have $p_T>20$ GeV and any additional lepton $p_T>10$ GeV. For Ak10 jets, the corresponding requirements are $p_T>200$ GeV for the leading jet and $p_T>150$ GeV for additional jets.

\section{Analysis Strategy}\label{analysis_strategy}

\begin{figure*}[tbp]
\centering
\begin{subfigure}{0.32\textwidth}
\includegraphics[width=1\textwidth]{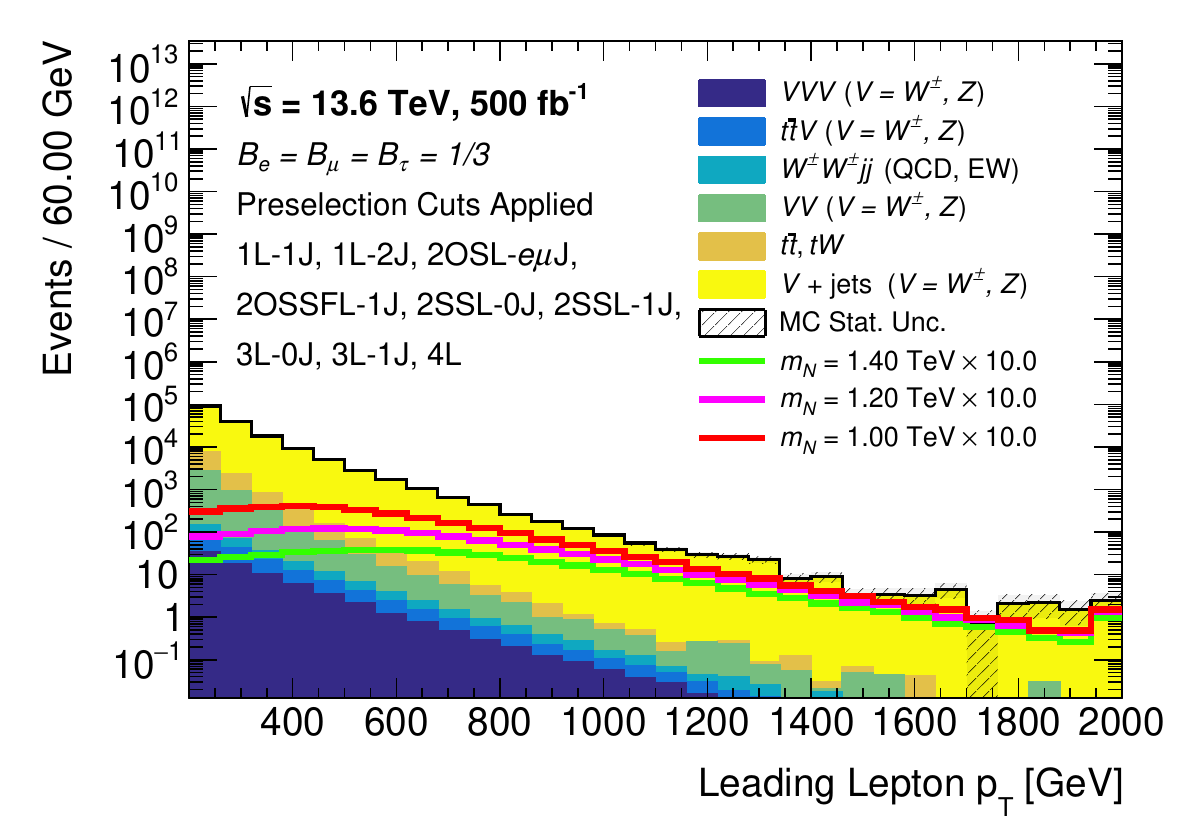}
\end{subfigure}
\hfill
\begin{subfigure}{0.32\textwidth}
\includegraphics[width=1\textwidth]{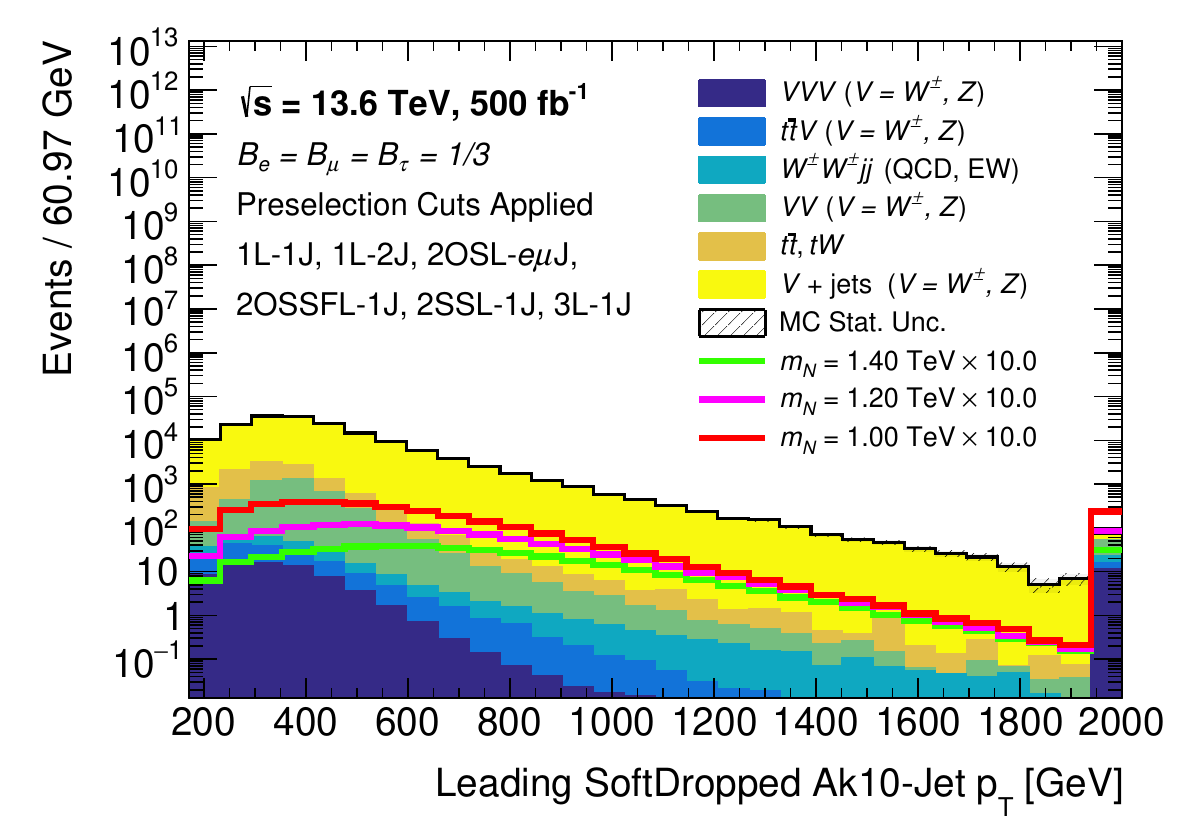}
\end{subfigure}
\hfill
\begin{subfigure}{0.32\textwidth}
\includegraphics[width=1\textwidth]{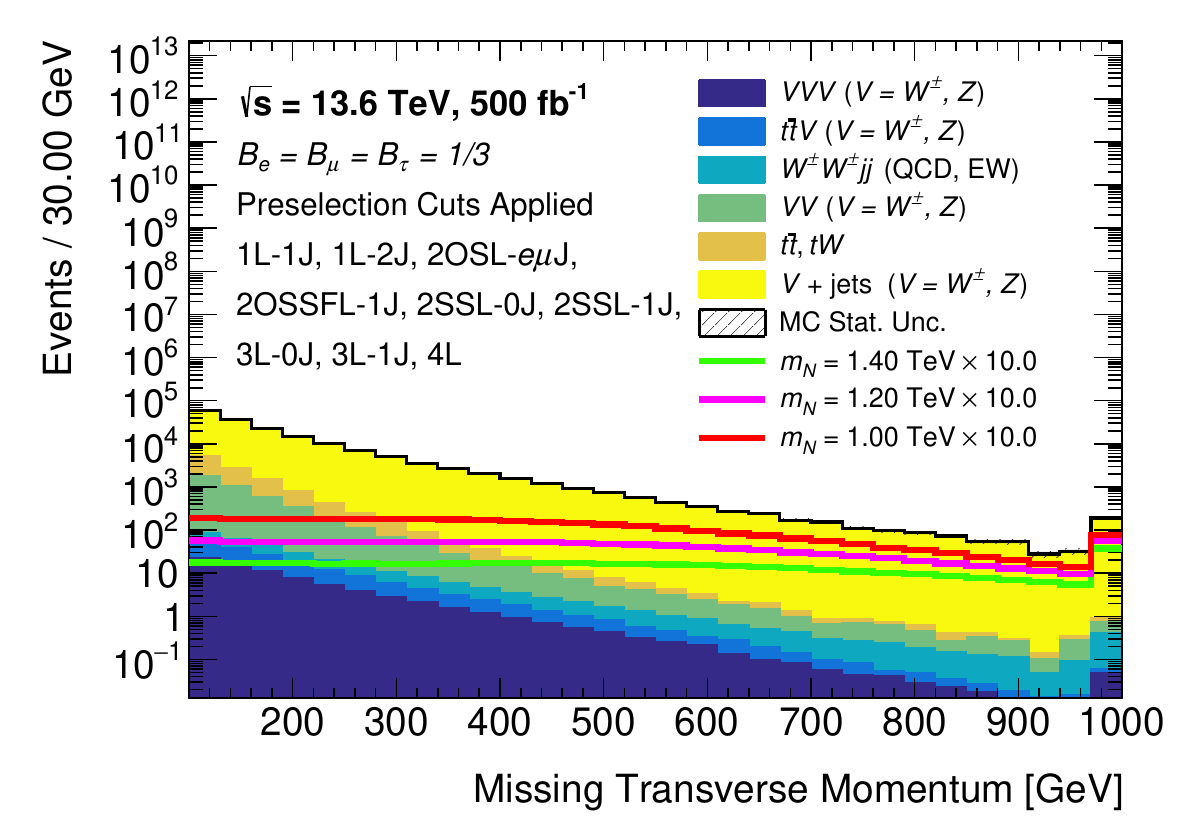}
\end{subfigure}
\hfill
\begin{subfigure}{0.325\textwidth}
\includegraphics[width=1\textwidth]{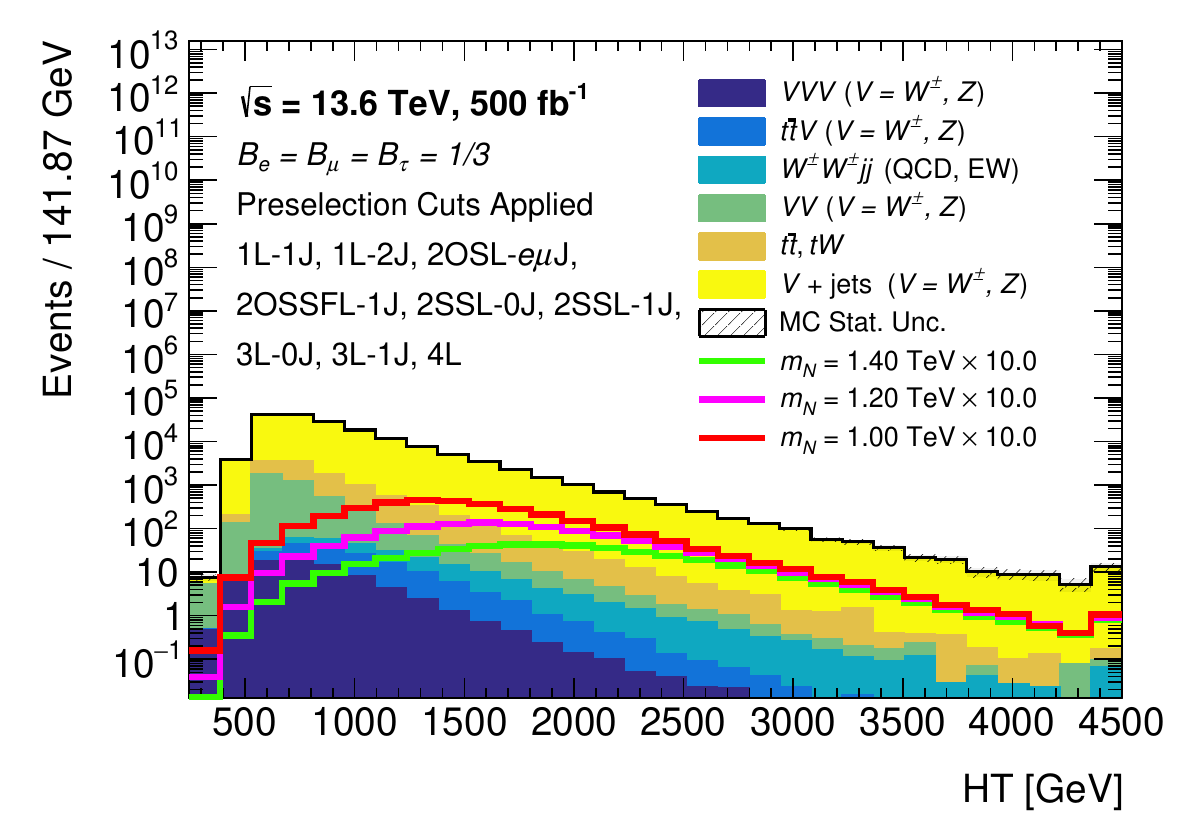}
\end{subfigure}
\hfill
\begin{subfigure}{0.325\textwidth}
\includegraphics[width=1\textwidth]{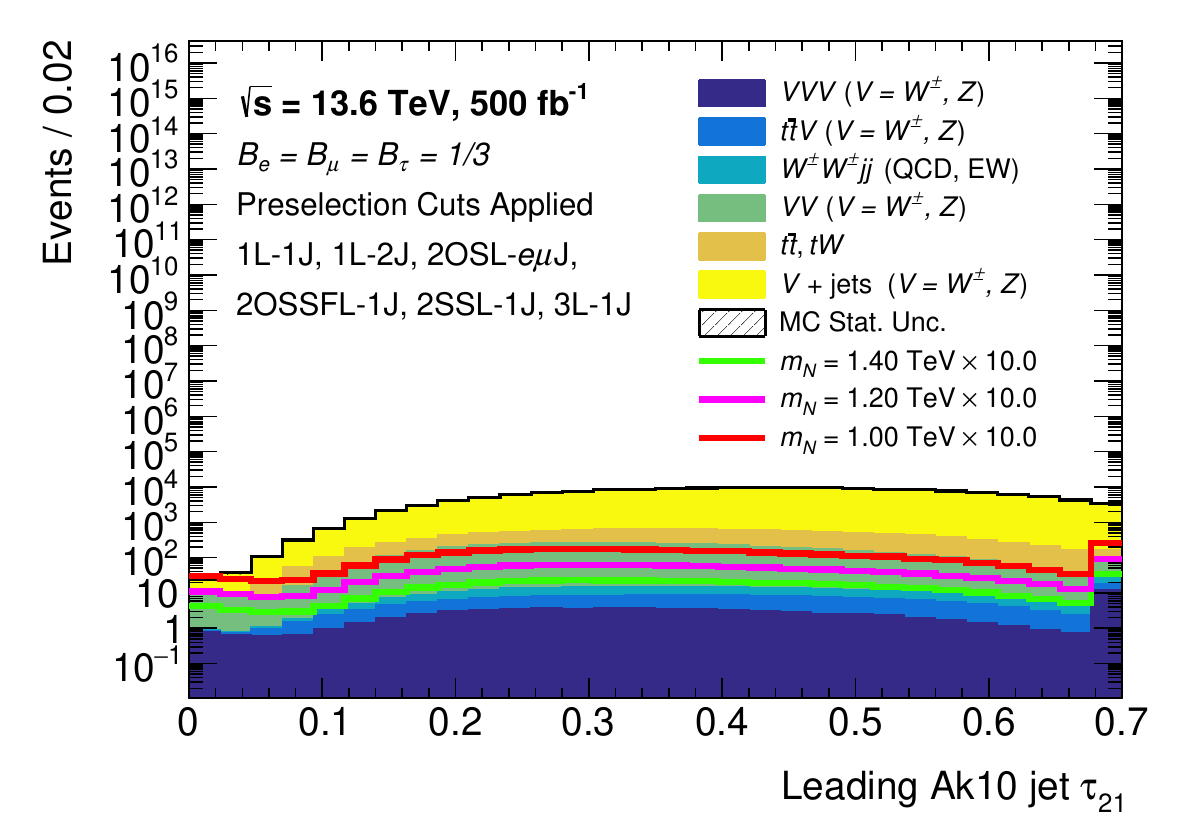}
\end{subfigure}
\hfill
\begin{subfigure}{0.325\textwidth}
\includegraphics[width=1\textwidth]{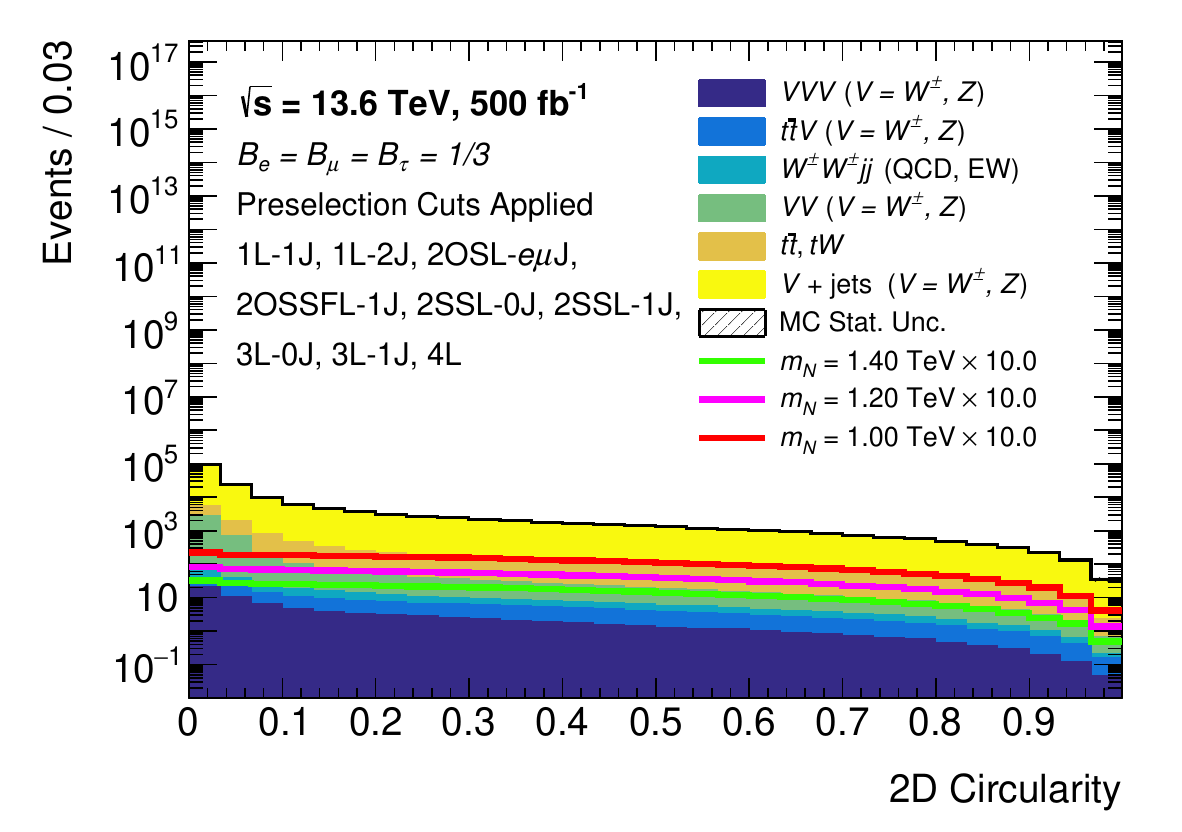}
\end{subfigure}
\caption{Distributions of the leading-lepton $p_T$ (upper left), leading soft-dropped Ak10 jet $p_T$ (upper middle), missing transverse momentum $p_T^{\rm miss}$ (upper right), hadronic transverse momentum sum $H_T$ (lower left), $\tau_{21}$ subjettiness ratio of the leading Ak10 jet (lower middle), and 2D circularity (lower right), in the six search channels containing at least one Ak10 jet of Section~\ref{analysis_strategy}. The rightmost (leftmost) bin contains the overflow (underflow) events in each distribution. The gray band represents the statistical uncertainty of the SM background prediction. As illustrative examples, signal hypothesis for the production of type-III seesaw heavy leptons in the FDM scenario with $m_N = 1.0$, $1.1$, and $1.2$~TeV are overlaid in all distributions.}
\label{feature_histograms_set1}
\end{figure*}

\begin{figure*}[ht]
\centering
\begin{subfigure}{0.32\textwidth}
\includegraphics[width=1\textwidth]{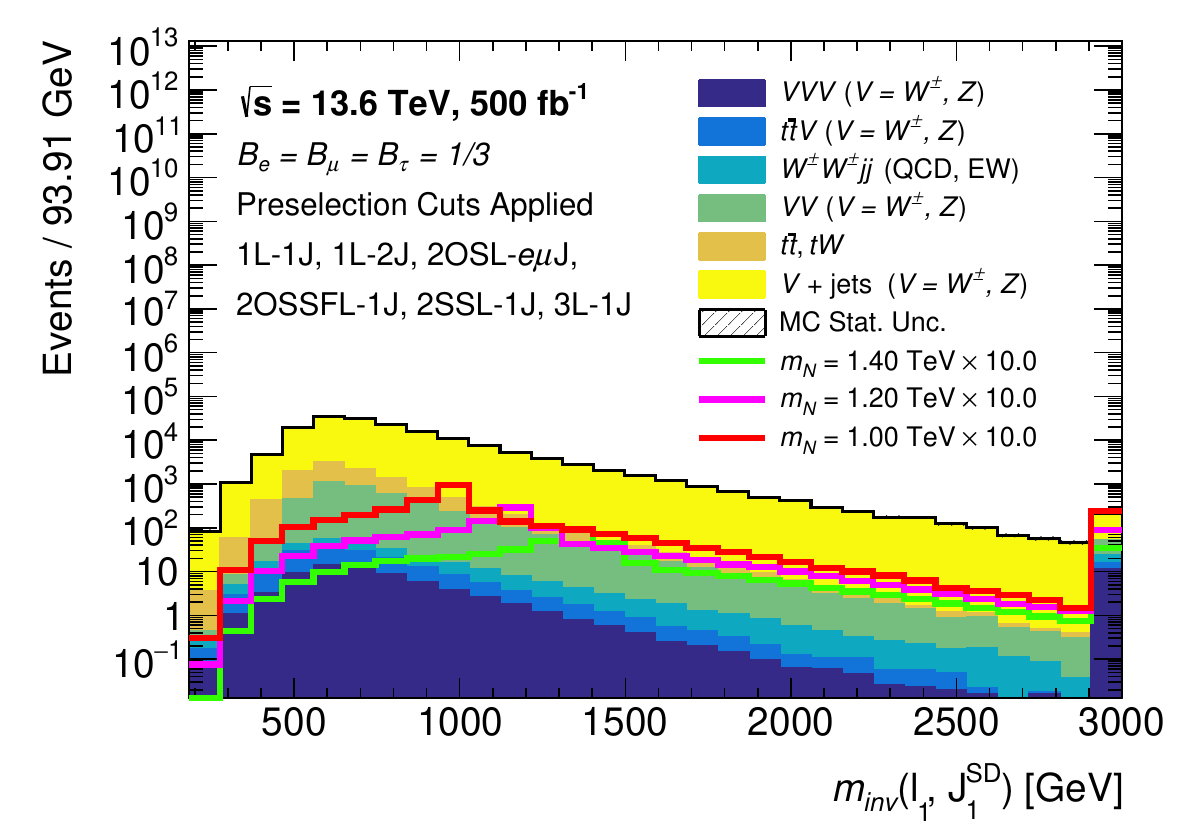}
\end{subfigure}
\hfill
\begin{subfigure}{0.32\textwidth}
\includegraphics[width=1\textwidth]{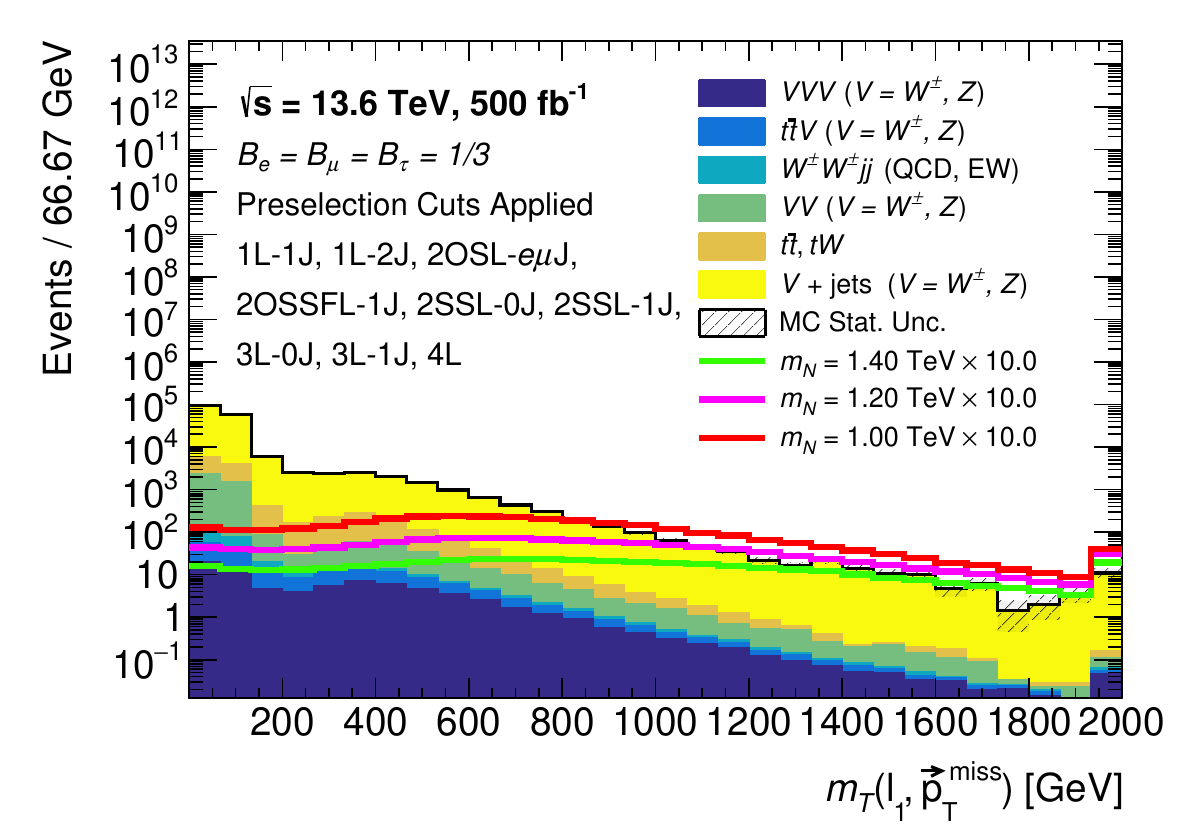}
\end{subfigure}
\hfill
\begin{subfigure}{0.32\textwidth}
\includegraphics[width=1\textwidth]{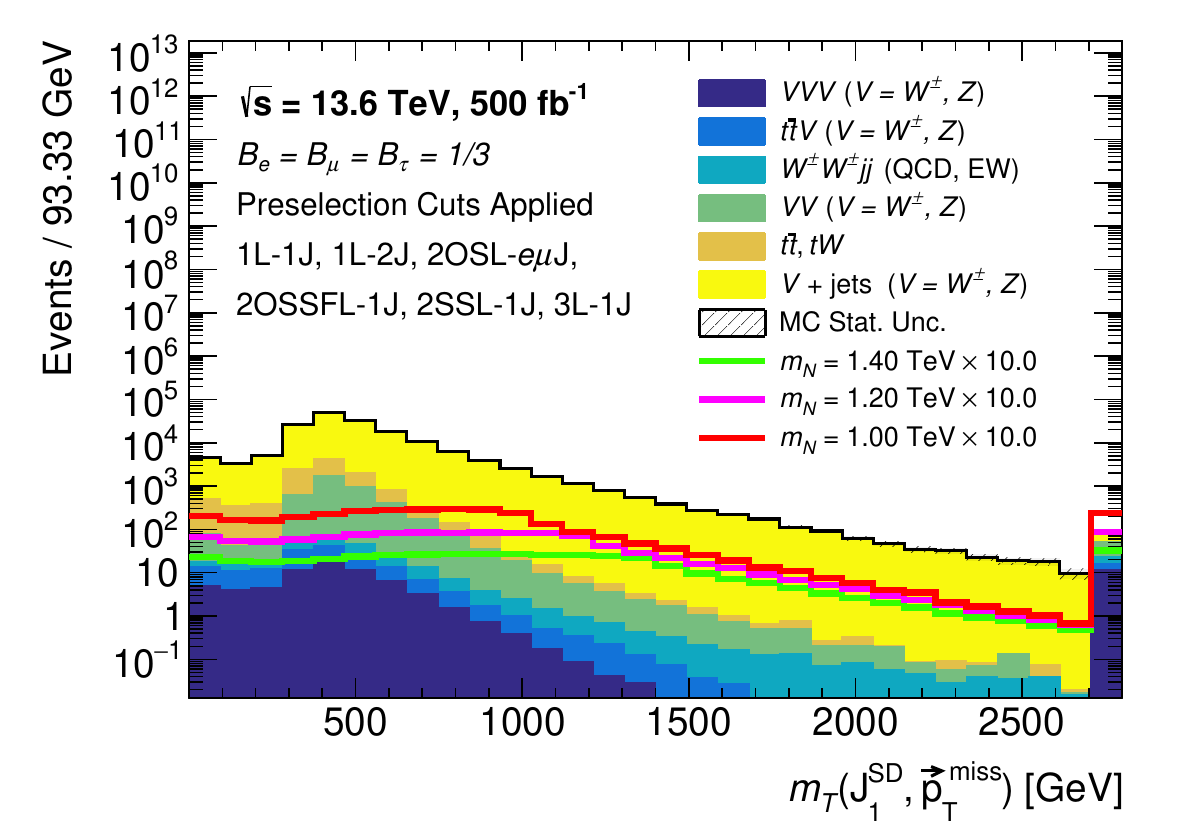}
\end{subfigure}
\hfill
\begin{subfigure}{0.325\textwidth}
\includegraphics[width=1\textwidth]{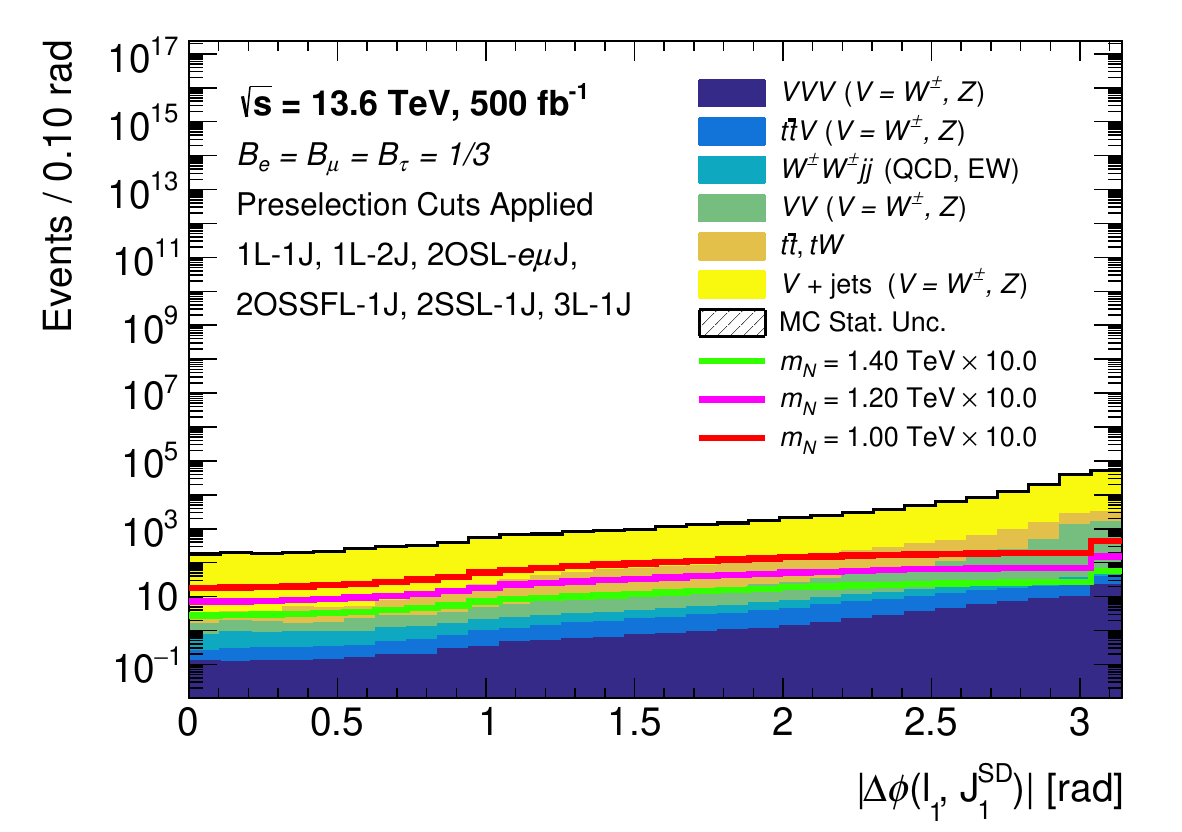}
\end{subfigure}
\hfill
\begin{subfigure}{0.325\textwidth}
\includegraphics[width=1\textwidth]{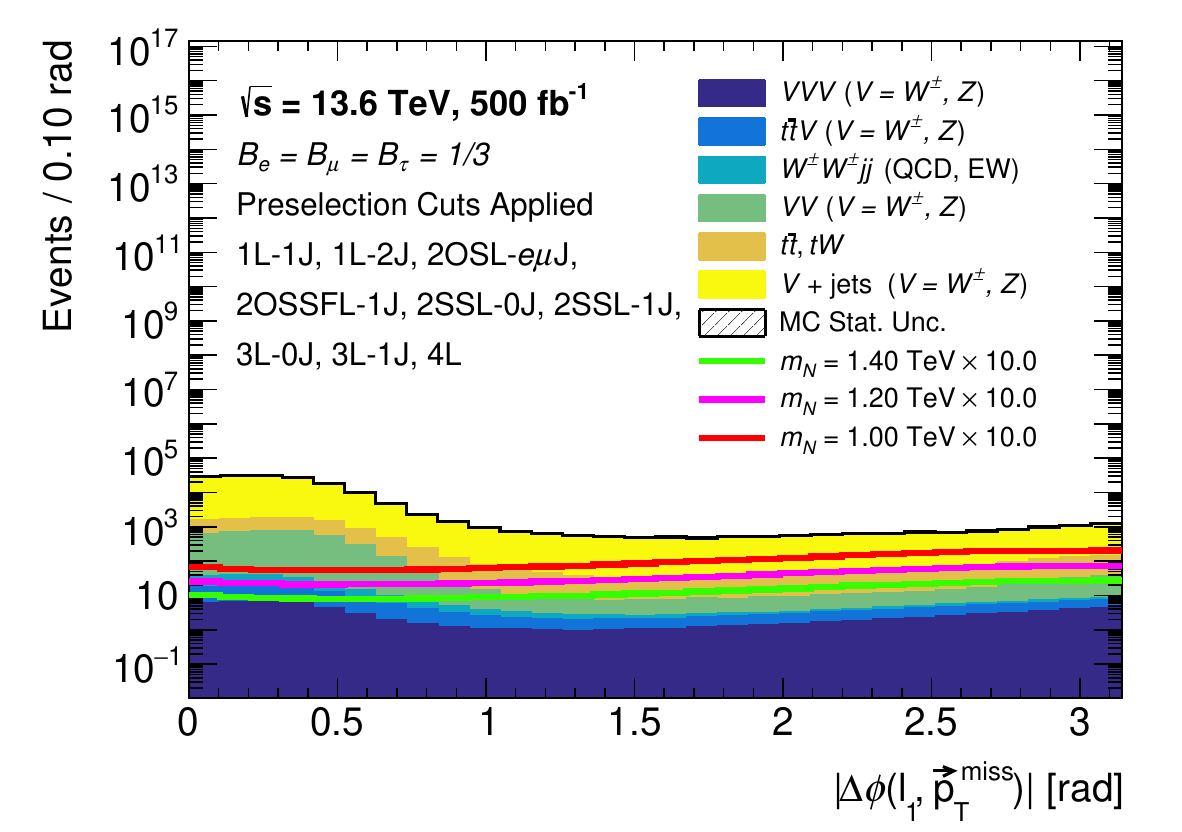}
\end{subfigure}
\hfill
\begin{subfigure}{0.325\textwidth}
\includegraphics[width=1\textwidth]{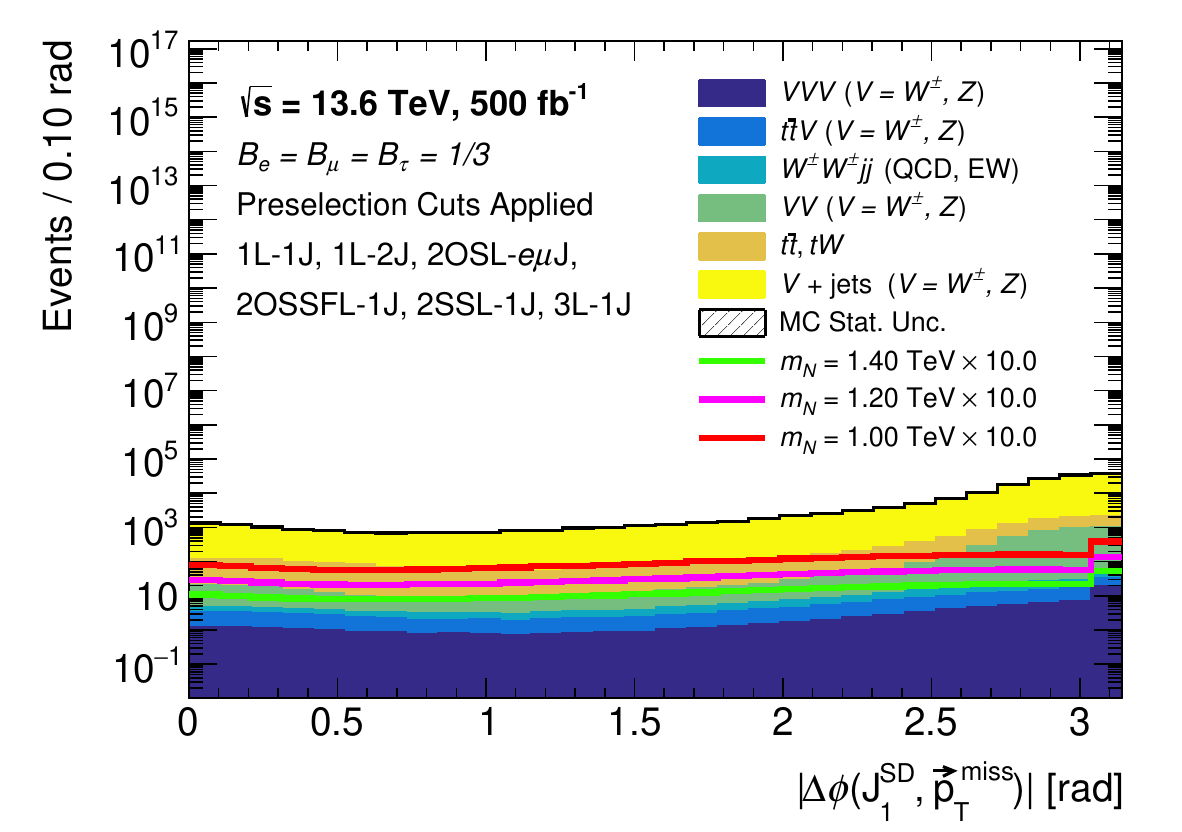}
\end{subfigure}
\caption{Distributions of the invariant mass $M_{\rm inv}(\ell_1, J_1^{\rm SD})$ (upper left), the transverse mass $m_T(\ell_1, \vec p_T^{\rm miss})$ (upper middle), $m_T(J_1^{\rm SD}, \vec p_T^{\rm miss})$ (upper right), and the absolute azimuthal separations $|\Delta\phi(\ell_1, J_1^{\rm SD})|$ (lower left), $|\Delta\phi(\ell_1, \vec p_T^{\rm miss})|$ (lower middle), and $|\Delta\phi(J_1^{\rm SD}, \vec p_T^{\rm miss})|$ (lower right), in the six search channels containing at least one Ak10 jet. The rightmost (leftmost) bin contains the overflow (underflow) events in each distribution. The gray band represents the statistical uncertainty of the SM background prediction. As illustrative examples, signal hypothesis for the production of type-III seesaw heavy leptons in the FDM scenario with $m_N = 1.0$, $1.1$, and $1.2$~TeV are overlaid in all distributions.}
\label{feature_histograms_set2}
\end{figure*}

\begin{figure*}[tbp]
\centering
\begin{subfigure}{0.329\textwidth}
\includegraphics[width=1\textwidth]{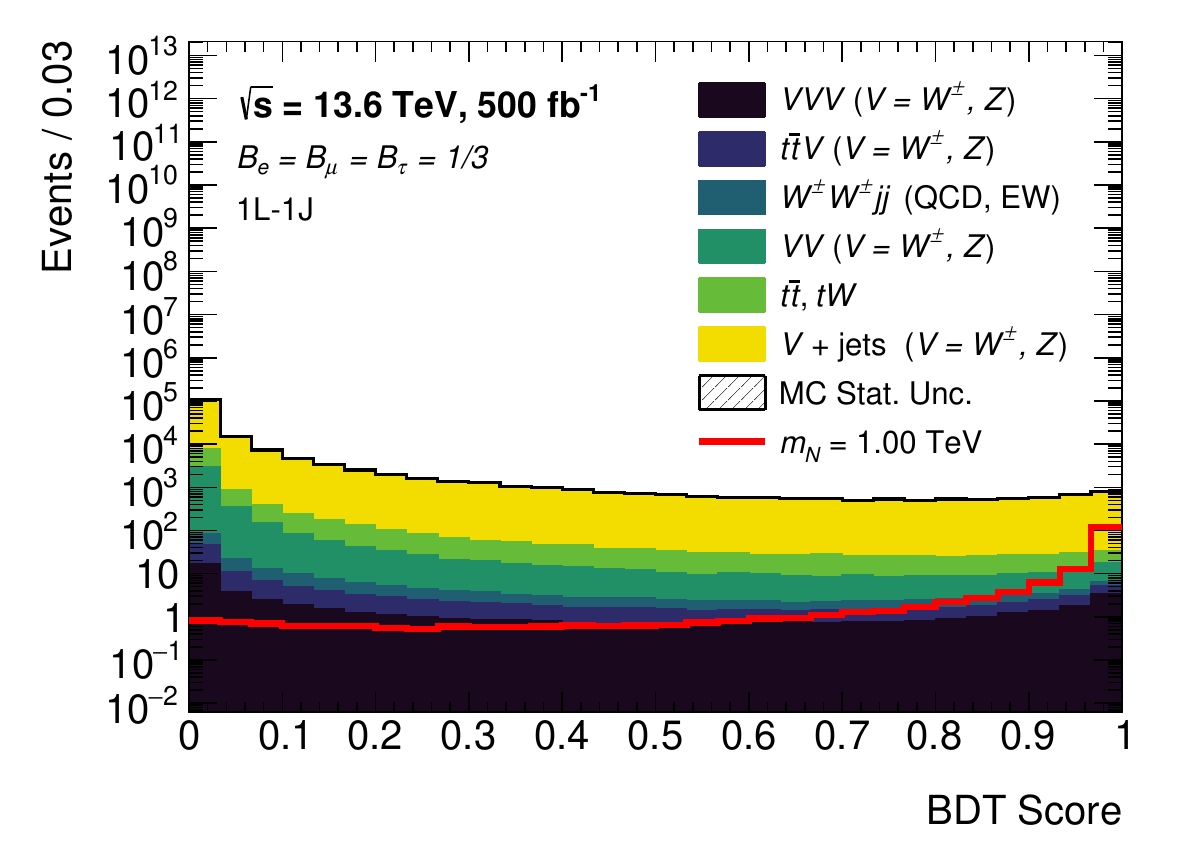}
\end{subfigure}
\hfill
\begin{subfigure}{0.329\textwidth}
\includegraphics[width=1\textwidth]{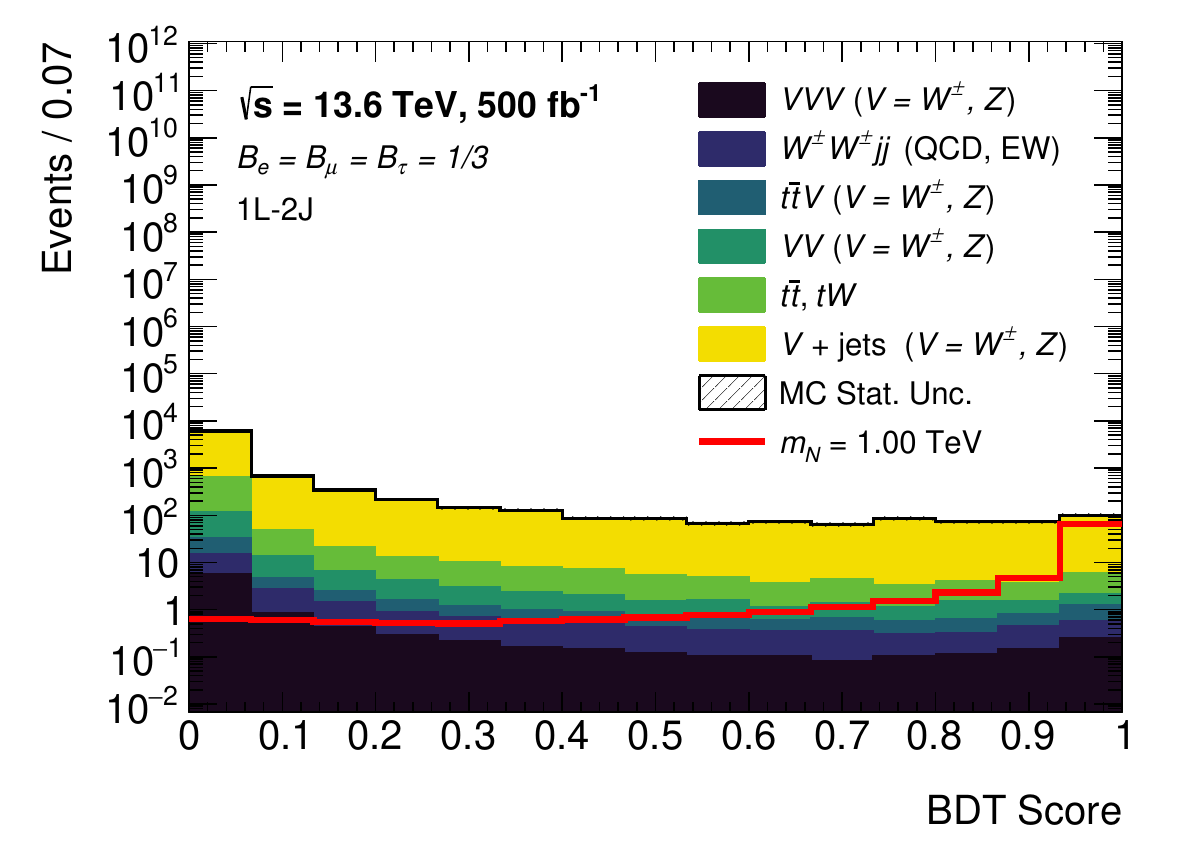}
\end{subfigure}
\hfill
\begin{subfigure}{0.329\textwidth}
\includegraphics[width=1\textwidth]{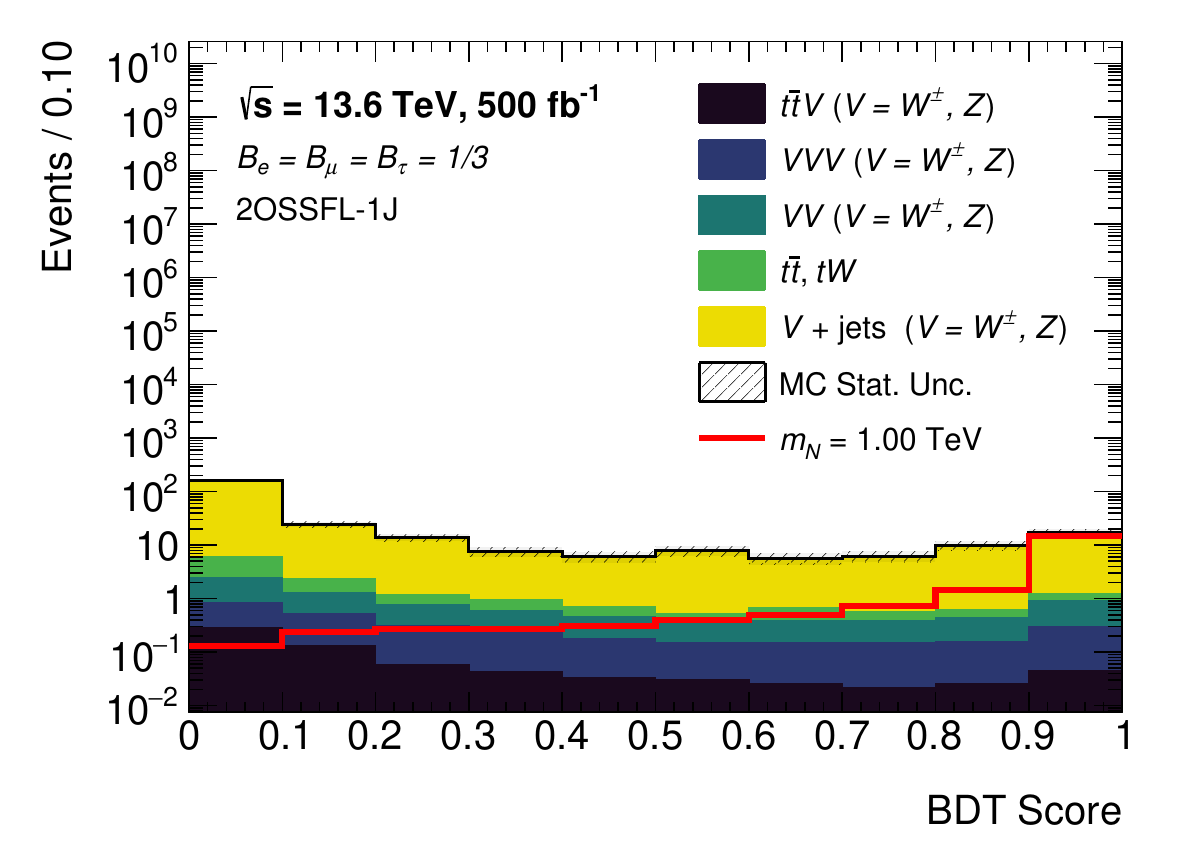}
\end{subfigure}
\hfill
\begin{subfigure}{0.329\textwidth}
\includegraphics[width=1\textwidth]{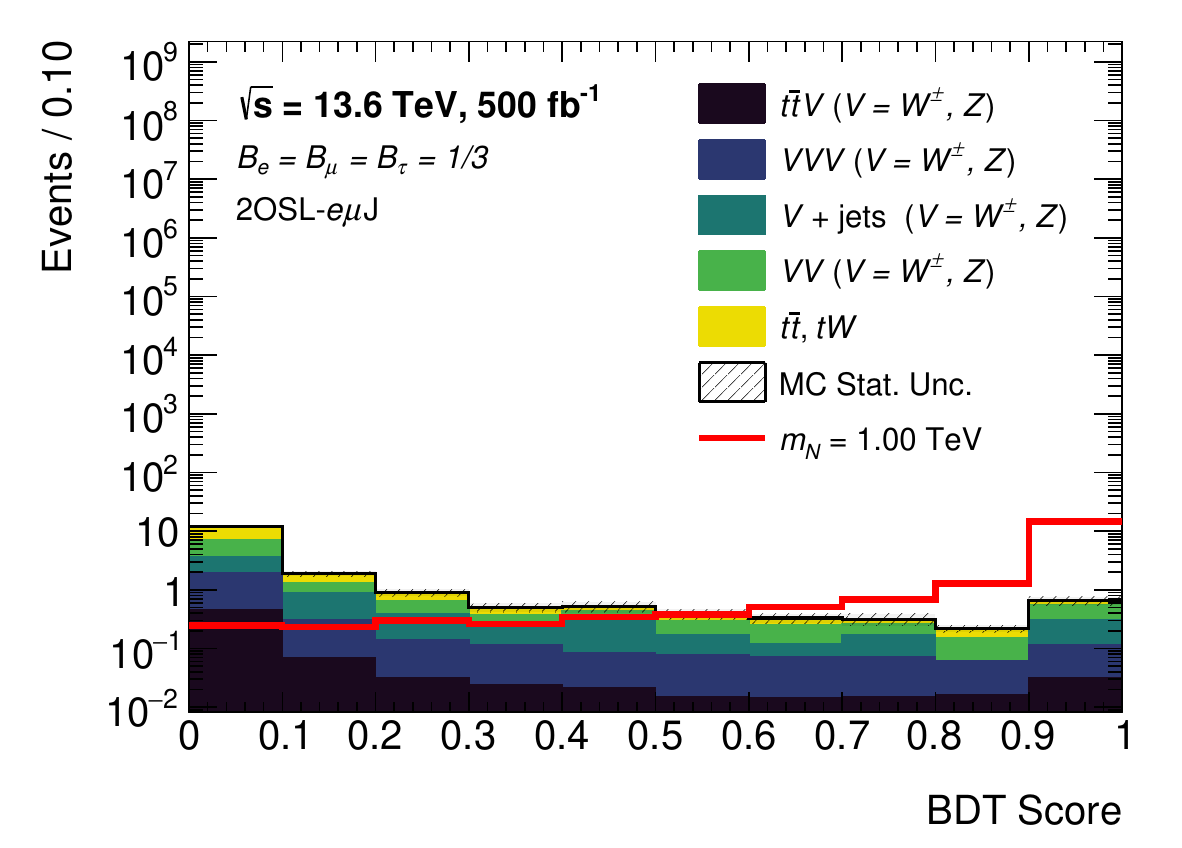}
\end{subfigure}
\hfill
\begin{subfigure}{0.329\textwidth}
\includegraphics[width=1\textwidth]{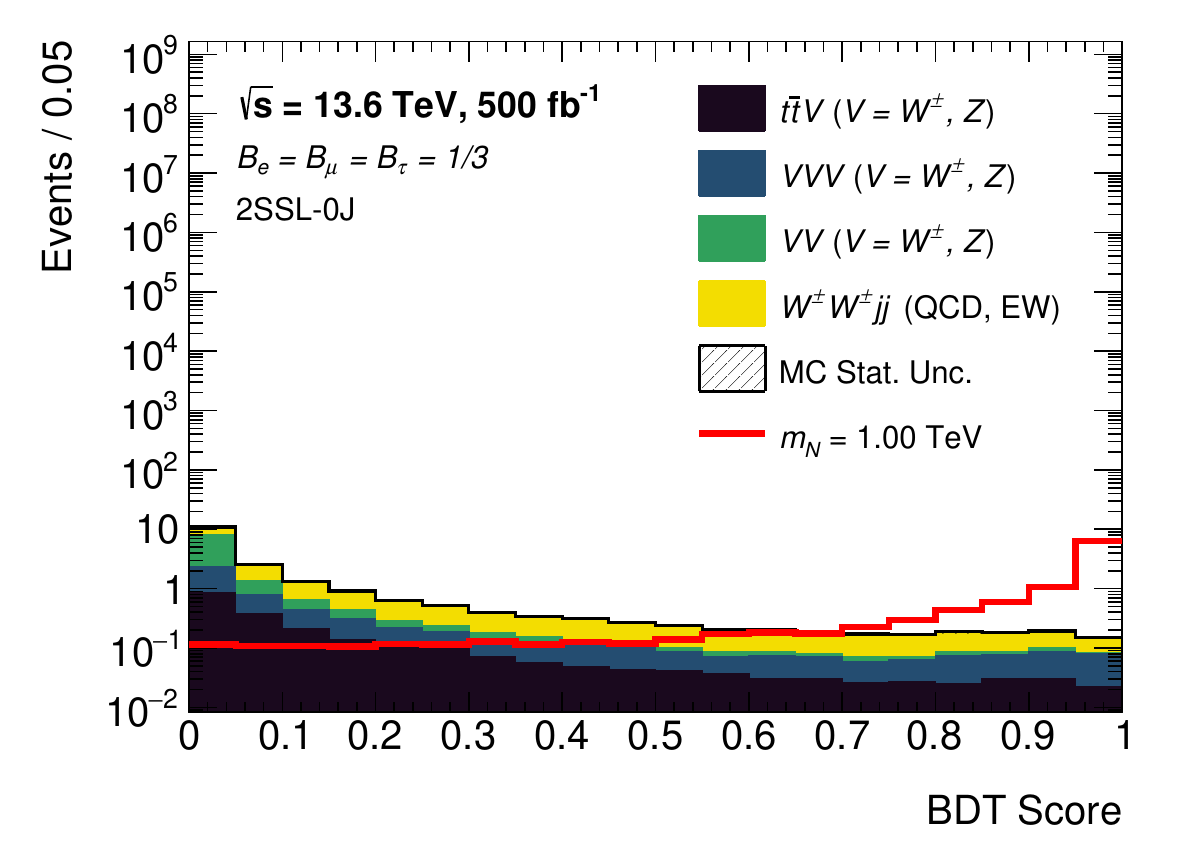}
\end{subfigure}
\hfill
\begin{subfigure}{0.329\textwidth}
\includegraphics[width=1\textwidth]{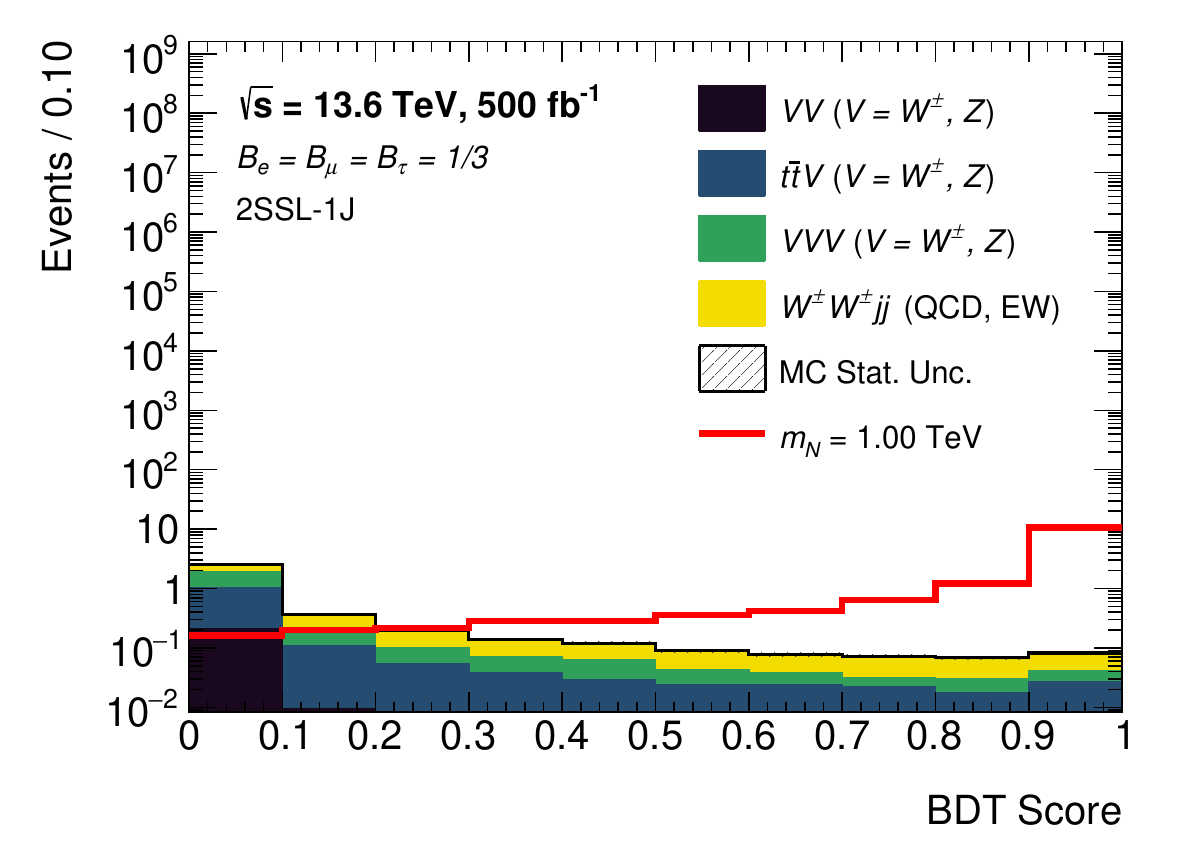}
\end{subfigure}
\hfill
\begin{subfigure}{0.329\textwidth}
\includegraphics[width=1\textwidth]{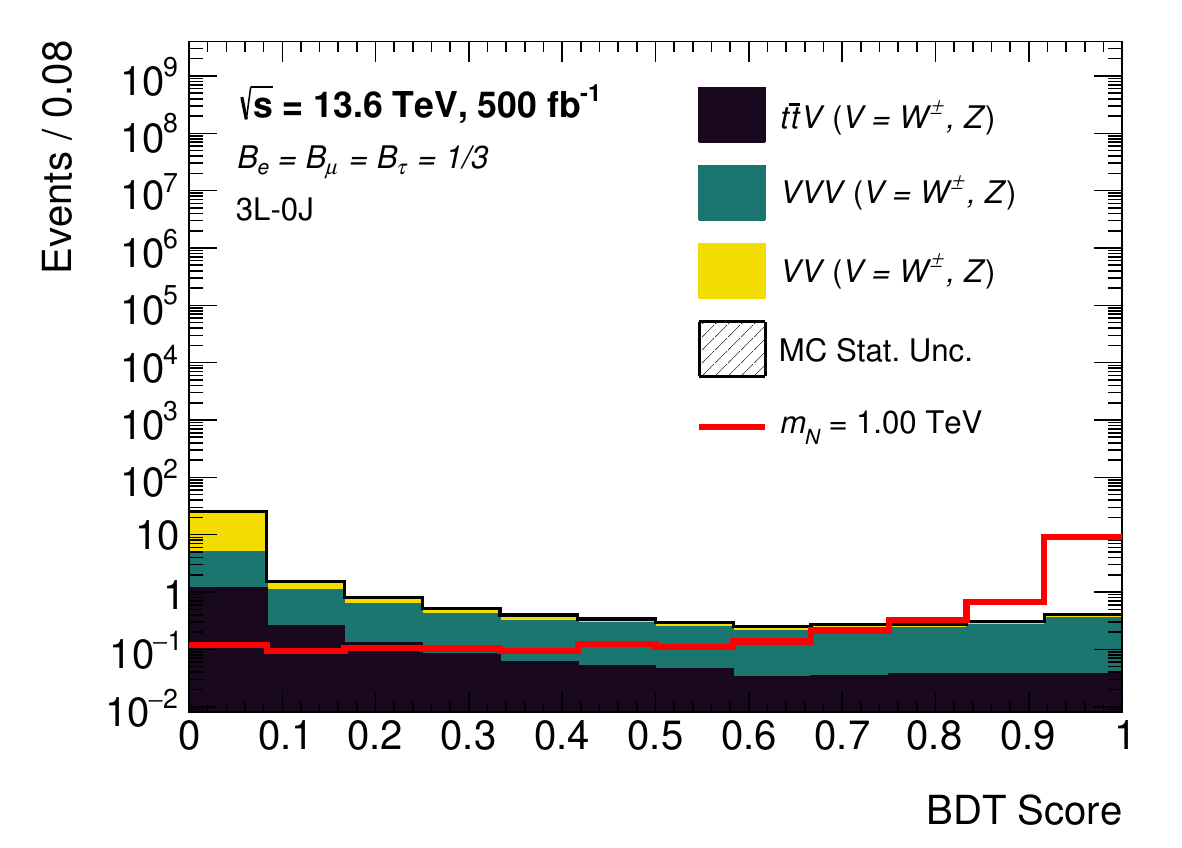}
\end{subfigure}
\hfill
\begin{subfigure}{0.329\textwidth}
\includegraphics[width=1\textwidth]{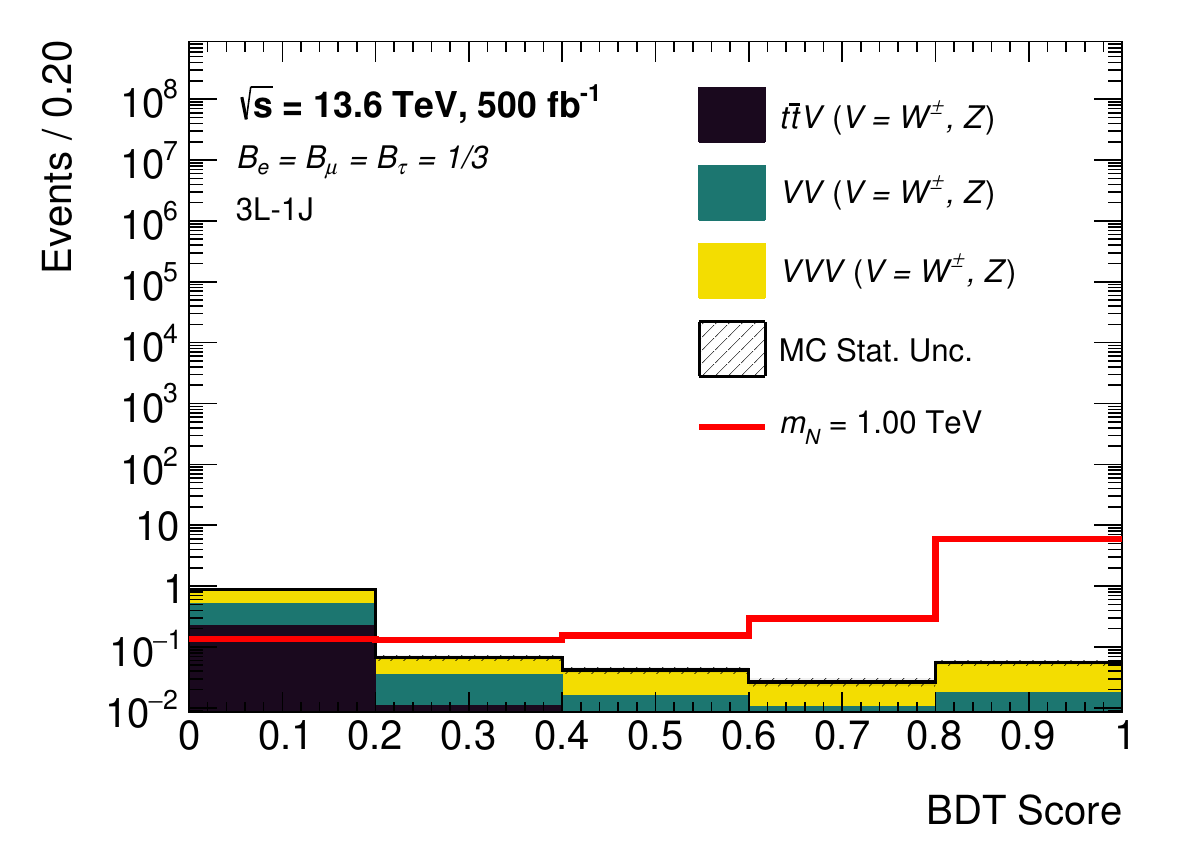}
\end{subfigure}
\hfill
\begin{subfigure}{0.329\textwidth}
\includegraphics[width=1\textwidth]{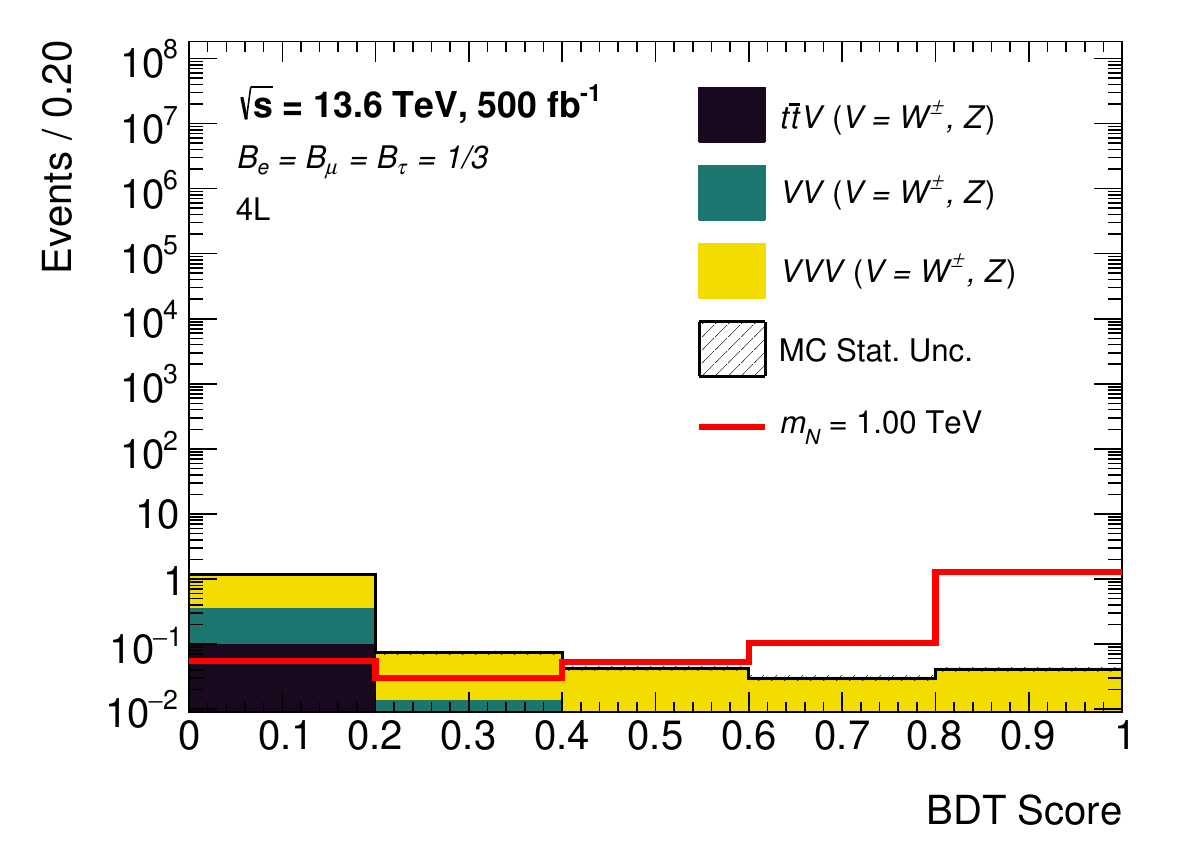}
\end{subfigure}
\caption{Distributions of the pBDT score in the 1L-1J, 1L-2J, 2OSSFL-1J, 2OSL-$e\mu$J, 2SSL-0J, 2SSL-1J, 3L-0J, 3L-1J and 4L search channels, arranged from upper left to lower right. The grey band represents the statistical uncertainty of the SM background prediction. An example signal hypothesis for $m_N = 1.0$~TeV is overlaid.}
\label{bdt_histograms}
\end{figure*}

Following object selection, events are classified into nine orthogonal channels based on the multiplicities of the selected leptons and Ak10 jets as described in Section~\ref{collider_signatures}. Events not falling into any of the categories are discarded. The common preselection requirements are then applied to all channels. The leading and subleading leptons, where present, are required to satisfy $p_T>200$ and $100$~GeV, respectively, and the missing transverse momentum must exceed $100$~GeV. To suppress backgrounds from low mass resonances and Drell-Yan processes, events containing an opposite sign same flavour (OSSF) lepton pair are required to satisfy $m_{\ell\ell}^\text{OSSF} > 12$ GeV and $\left| m_{\ell\ell}^\text{OSSF} - m_Z \right| > 15$ GeV, respectively. The $Z$-veto is not applied in the 3L and 4L channels to avoid rejecting signal events containing leptonic $Z$ decays. No channel-specific cuts were further applied to keep sufficient amount of events for training and testing the multivariate classifiers. The expected signal and background event yields after all selection criteria are shown in Table~\ref{channel_yields}.

Figures~\ref{feature_histograms_set1} and \ref{feature_histograms_set2} display the kinematic distributions of the discriminating observables, combined across all channels after the preselection. Figure~\ref{feature_histograms_set1} shows the transverse momentum distributions of the leading lepton, the leading soft-dropped Ak10 jet, and the missing transverse momentum (first row), followed by the scalar sum of hadronic transverse momenta ($H_T$), $\tau_{21}$ subjettiness ratio of the leading Ak10 jet (lower middle), and 2D circularity (second row). Figure~\ref{feature_histograms_set2} displays high-level event features. The first row shows the invariant mass of the leading lepton and the leading soft-dropped Ak10 jet, $m_\text{inv}\left(\ell_1, J_1^\text{SD}\right)$, alongside the transverse masses $m_T\left(\ell_1, \vec{p}_T^\mathrm{~miss}\right)$ and $m_T\left(J_1^\text{SD}, \vec{p}_T^\mathrm{~miss}\right)$, where the transverse mass is defined as $m_T\left(\mathcal{O}, \vec{p}_T^{\text{ miss}}\right) = \sqrt{ 2\left|\vec{p}_T^{\text{ miss}}\right|\left|\vec{p}_T^{\text{ $\mathcal{O}$}}\right|\left\{1-\cos\left[\Delta\phi\left(\vec{p}_T^{\text{ miss}}, \vec{p}_T^{\text{ $\mathcal{O}$}}\right)\right]\right\} }$. The lower panel shows the absolute azimuthal separations between the leading lepton, the leading soft-dropped jet, and the missing transverse momentum. In all distributions, the signal exhibits harder kinematics than the background, as expected for the TeV-scale heavy leptons considered here. The discriminating power of these observables follows directly from the signal topology. When the selected lepton and the reconstructed boosted boson originate from the same heavy-lepton decay chain, the invariant mass $m_\text{inv}\left(\ell_1, J_1^\text{SD}\right)$ peaks above the truth signal mass $m_N$. The transverse-mass distribution $m_T\left(J_1^\text{SD}, \vec{p}_T^\mathrm{~miss}\right)$ exhibits a Jacobian-like feature near $m_N$, arising primarily from the $N^\pm \to W^\pm_\text{had} \nu$ decay mode. Finally, the azimuthal separations $|\Delta\phi(\ell_1, \vec{p}_T^\mathrm{~miss})|$, $|\Delta\phi(J_1^\text{SD}, \vec{p}_T^\mathrm{~miss})|$, and $|\Delta\phi(\ell_1, J_1^\text{SD})|$ encode correlations among objects produced in the heavy lepton pair decays.

\begin{table}[tbp]
\small
\centering
\setlength{\tabcolsep}{10pt}
\renewcommand{\arraystretch}{1.5}
\begin{tabular}{lll}
\toprule
Channel & $\sum w_s \pm \sqrt{\sum w^2_s}$ & $\sum w_b \pm \sqrt{\sum w^2_b}$ \\
\midrule
1L-1J & $166.17 \pm 0.45$ & $161674.48 \pm 315.18$ \\
1L-2J & $82.07 \pm 0.32$ & $8330.39 \pm 70.45$ \\
2OSSFL-1J & $18.91 \pm 0.15$ & $259.24 \pm 9.56$ \\
2OSL-e$\mu$J & $18.80 \pm 0.15$ & $17.88 \pm 0.50$ \\
2SSL-0J & $10.68 \pm 0.11$ & $20.01 \pm 0.10$ \\
2SSL-1J & $14.45 \pm 0.13$ & $3.78 \pm 0.04$ \\
3L-0J & $11.21 \pm 0.12$ & $31.17 \pm 0.16$ \\
3L-1J & $6.69 \pm 0.09$ & $1.08 \pm 0.02$ \\
4L & $1.53 \pm 0.04$ & $1.36 \pm 0.02$ \\
\bottomrule
\end{tabular}
\caption{Sum of signal weights and background weights in each
analysis channel after object selection, channel classification
and preselection. The results are shown for the
$m_N=1.0$~TeV signal mass hypothesis. The statistical
uncertainties arise from the finite sizes of the simulated samples.}
\label{channel_yields}
\end{table}

\begin{table}[tbp]
\small
\setlength{\tabcolsep}{6pt}
\renewcommand{\arraystretch}{1.5}
\begin{tabular}{lll}
\toprule
Object(s) & Variables & Channels \\
\midrule
Lepton & $p_T$ & All \\
Ak10 jet ($\Tilde{J}$) & $p_T$, $\tau_{21}$, $\tau_{32}$ & At least 1J \\
$\ell$ -- $\Tilde{J}$ pair & $m_\mathrm{inv}$, $|\Delta\phi|$ & At least 1J  \\
$\ell$ -- $\ell$ pair & $m_\mathrm{inv}$, $|\Delta\phi|$ & 2OSL, 2SSL, 3L, 4L \\
$\Tilde{J}$ -- $\Tilde{J}$ pair & $m_\mathrm{inv}$, $|\Delta\phi|$ & 1L-2J \\
$\vec{p}_T^{\text{ miss}}$ & $\left|\vec{p}_T^{\text{ miss}}\right|$ & All \\
$\ell$ -- $\vec{p}_T^{\text{ miss}}$ & $m_T$, $|\Delta\phi|$ & All \\
$\Tilde{J}$ -- $\vec{p}_T^{\text{ miss}}$ & $m_T$, $|\Delta\phi|$ & At least 1J  \\
Global & $H_T$, $\mathcal{S}(E_T^\text{miss})$, E.S.  & All \\
\bottomrule
\end{tabular}
\caption{Input variables used to train the \textsc{XGBoost} classifiers, grouped by the object(s) from which they are constructed. Here $\ell$ denotes the selected lepton and $\Tilde{J}$ denotes the selected soft-dropped Ak10 jet. Wherever $|\Delta\phi|$ is included, only the minimum and maximum across all possible pairs is included. By E.S. we refer to event shapes. Channels that contain at least one Ak10 are 1L, 2OSL, 2SSL-1J, 3L-1J (six channels).}
\label{bdt_variables}
\end{table}

To enhance the search sensitivity, multivariate analysis techniques are employed to separate the signal events from the remaining SM background events using mass-parameterized boosted decision trees (pBDT) \cite{pbdt}.  The mass parameterization is implemented by stacking all the $N$ signal samples into a single sample and supplying the true signal mass $m_N$ as an additional input feature. For the background events, an oversampling method is used rather than randomly assigning $m_N$ values. Each background event is replicated $N$ times in the training sample, with a different signal mass value assigned to each copy. To avoid biasing the training towards a specific signal mass hypothesis, the signal samples are normalized such that the sum of weights is the same for all mass points. Furthermore, all of the signal mass hypotheses are collectively normalized so that their sum of weights is the same as the one for the background samples, forcing the classifier to discriminate based on shape differences between signal and background. The mass-parameterization approach allows the classifier to learn a mass-dependent decision function and the model can be evaluated at signal mass hypotheses not included in the training sample without retraining.

The input variables used to train the classifiers are summarized in Table~\ref{bdt_variables}. They are constructed from the selected leptons, soft-dropped Ak10 jets (denoted as $\Tilde{J}$), and the missing transverse momentum vector. Each signal region employs the subset of variables appropriate to its reconstructed final state. Individual-object observables include the lepton $p_T$ and the Ak10 jet $p_T$, and $N$-subjettiness variables $\tau_{21}$, and $\tau_{32}$. These variables characterize the internal jet structure and discriminate between boosted hadronic boson decays and QCD jets. Correlations between reconstructed objects are captured by the invariant masses $m_{\rm inv}$ and $\min|\Delta\phi|$, $\max|\Delta\phi|$ for all $(\ell, \Tilde{J})$ and $(\ell, \ell)$ combinations. Variables involving missing transverse momentum include the missing transverse energy (MET) $|\vec{p}_T^{\rm miss}|$ and the transverse masses $m_T(\ell,\vec{p}_T^{\rm miss})$ and $m_T(\Tilde{J},\vec{p}_T^{\rm miss})$. Finally, global event observables characterize the event topology, including the sum of hadronic transverse momenta $H_T$ and the MET signfiance $\mathcal{S}(E_T^\text{miss})$. Event shapes such as sphericity, aplanarity, circularity and centrality \cite{event_shapes} calculated using the combined four-momenta of all the selected leptons, Ak10 jets and MET are considered as well.

To evaluate the pBDT performance without bias from training, a two-fold out-of-fold (OOF) cross-training strategy is employed. The full dataset is split into two statistically disjoint halves, denoted \textsc{DS1} and \textsc{DS2}. \textsc{Model 1} is trained on \textsc{DS1} and used to generate predictions for \textsc{DS2}, while \textsc{Model 2} is trained on \textsc{DS2} and used to generate predictions for \textsc{DS1}. This ensures that the score distributions presented in the final analysis are derived from events not used during training. Moreover, this allows us to use all the generated events for performing the analysis, thus reducing the statistical uncertainty due to the limited MC sample sizes. During the training of each model, 20\% of the respective training half is reserved as a validation set. Both models use the same hyperparameters and are trained with the \textsc{XGBoost} algorithm \cite{xgboost}. Each tree is constrained to a maximum depth of 5 and grown with a learning rate of 0.05. The ensemble is grown for up to 3000 boosting rounds, but to control overtraining early stopping is applied. Early stopping terminates training if the validation-set logistic loss does not improve over 20 consecutive trees. The training weights are normalized to unit mean, so that the regularization parameters (L1, L2 penalties and $\gamma$ \cite{xgboost}) are kept at their library defaults, while the stochastic subsampling fractions for rows and features \cite{FRIEDMAN2002} are fixed to $80\%$. In addition, the minimum sum of Hessian weights in a leaf is not tuned but derived from the effective sample size of the training weights, ensuring that each leaf is supported by a minimum number of effective simulated events. The Kolmogorov-Smirnov test is used as an additional diagnostic of overtraining, with $p$-values above $0.05$ taken to indicate no statistically significant difference between the compared score distributions As shown in Figure \ref{bdt_histograms}, the pBDTs discriminate very well between the signal and background events, separating them apart across the score range.

\section{Results and Discussion}\label{results}

\begin{figure*}[ht!]
\centering
\begin{subfigure}{0.48\textwidth}
\centering
\includegraphics[width=\textwidth]{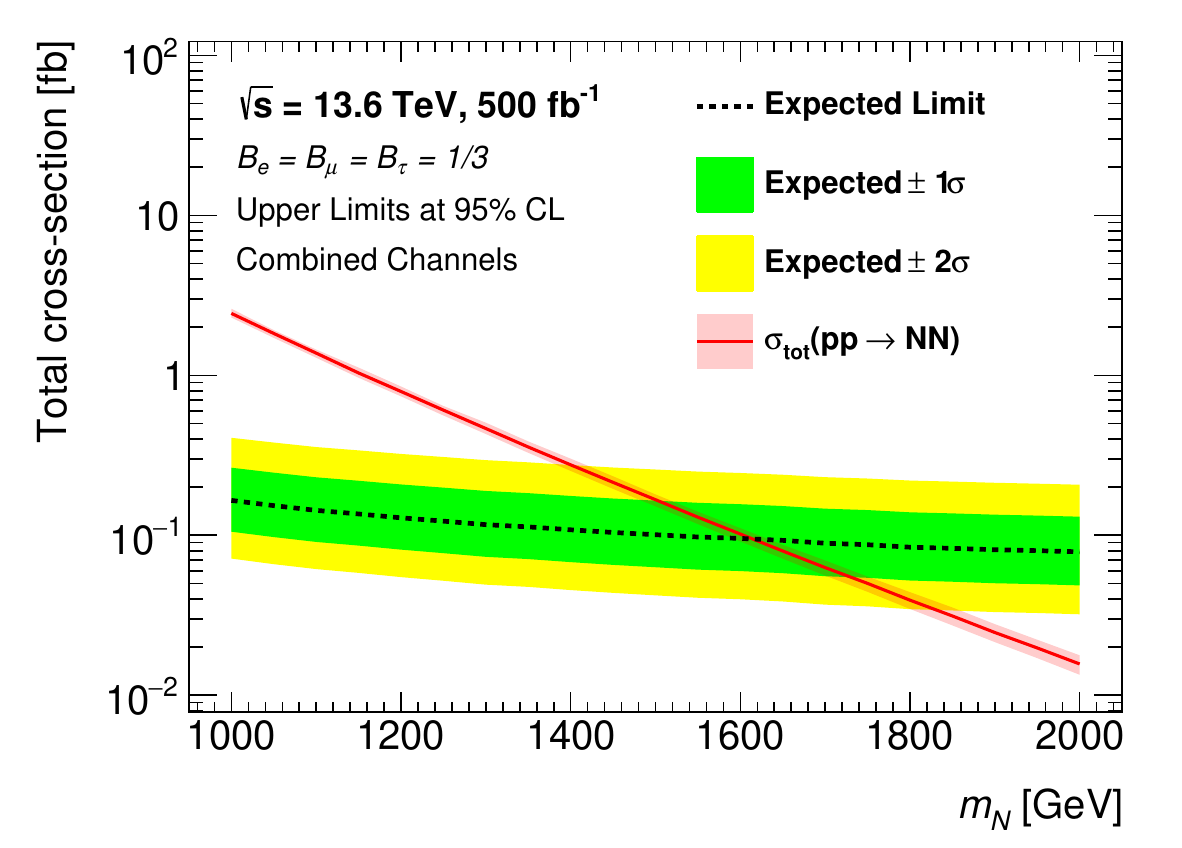}
\caption{Expected upper limits on the total cross section.}
\label{upper_limits_combined}
\end{subfigure}
\hfill
\begin{subfigure}{0.48\textwidth}
\centering
\includegraphics[width=\textwidth]{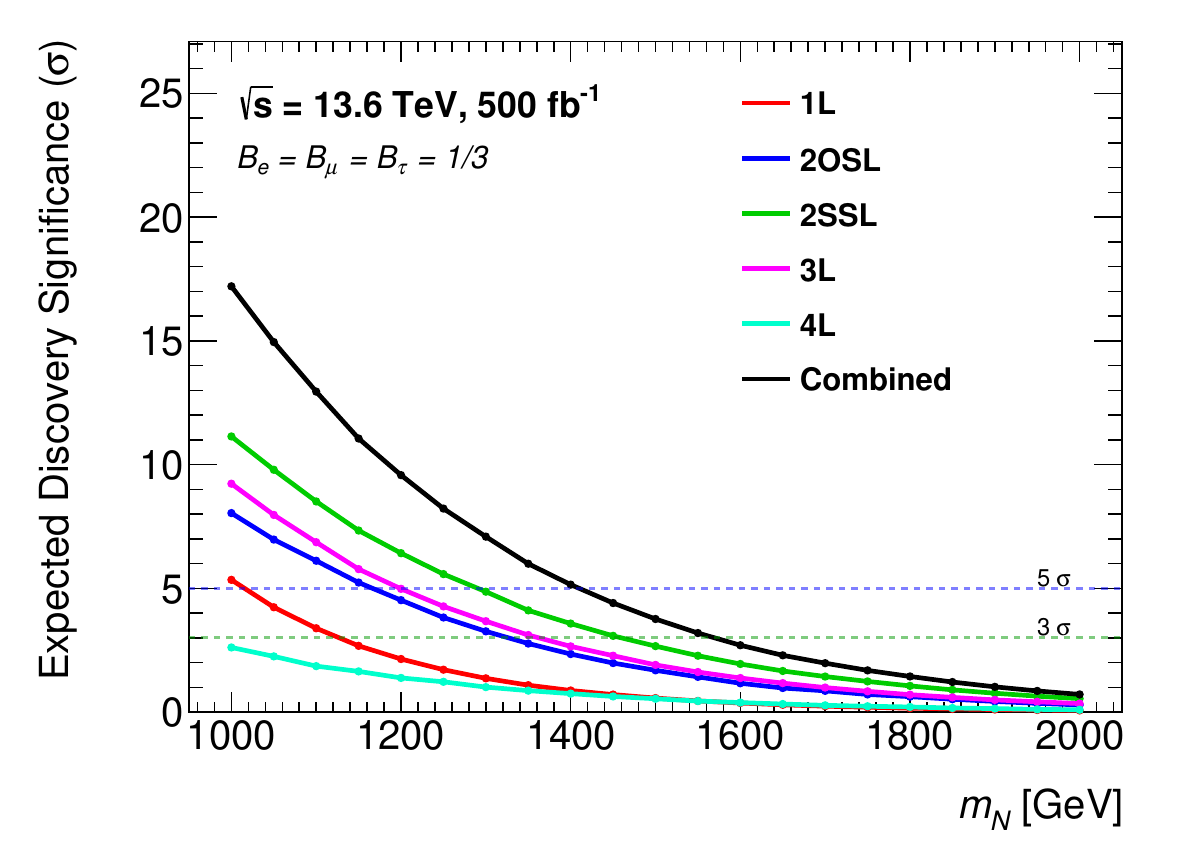}
\caption{Expected discovery significance.}
\label{significance}
\end{subfigure}
\caption{Expected $95\%$ CL upper limits on the signal production cross section for type-III seesaw heavy leptons in the FDM scenario, derived from the combined pBDT score distributions of all search channels (left), and expected discovery significance for the individual search channels and their combination (right).}
\label{results_plot}
\end{figure*}

The statistical analysis is performed using the \textsc{HistFactory}~\cite{histfactory} package, which implements a binned profile-likelihood fit to the pBDT score distributions. For each signal mass hypothesis, one-dimensional binned templates are constructed for the signal and total SM background in each channel. To balance the shape sensitivity of the classifier output against statistical fluctuations arising from the finite simulated samples, the number of bins is chosen separately for each channel: 30 for 1L-1J, 15 for 1L-2J, 10 for each of the two 2OSL channels and 2SSL-1J, 20 for 2SSL-0J, 12 for 3L-0J, and 5 for each of 3L-1J and 4L. This channel-dependent binning is chosen to retain the shape information while maintaining adequate background statistics and reducing fluctuations in sparsely populated bins. A flat $20\%$ normalization uncertainty is assigned to the total background in each channel to account for combined experimental and theoretical effects, excluding the luminosity uncertainty. This uncertainty is treated as an independent nuisance parameter for each channel. A flat $2\%$ luminosity uncertainty is applied to all simulated samples through a single correlated nuisance parameter. The statistical uncertainties arising from the finite size of the background samples are incorporated using per-bin nuisance parameters with Poisson constraints. Expected results are evaluated in the asymptotic approximation~\cite{cowan2011}. The expected discovery significance is derived from the profile-likelihood-ratio test statistic for the background-only hypothesis, evaluated using an Asimov dataset generated under the signal-plus-background hypothesis. The expected $95\%$ confidence level (CL) upper limits on the signal production cross section are obtained using the $\mathrm{CL}_s$ criterion~\cite{cls}, with a background-only Asimov dataset. Since all search channels are mutually exclusive, their likelihoods are multiplied to construct a simultaneous fit with a common signal-strength parameter $\mu$. The results are shown in Figures~\ref{significance} and~\ref{upper_limits_combined}. The combined pBDT fit yields an expected $95\%$ CL mass-exclusion reach of $1613~\mathrm{GeV}$ in the FDM scenario for the assumed $500~\mathrm{fb}^{-1}$ benchmark. The expected $5\sigma$ ($3\sigma$) discovery reach of the combined channels is $1410$ ($1570$)~GeV.

To assess the robustness of the expected exclusion sensitivity against background normalization and luminosity uncertainties, an additional statistical test is performed by setting the background normalization uncertainty to $50\%$ and the luminosity uncertainty to $2\%$, and comparing the result with an idealized reference scenario in which both uncertainties are set to zero. Relative to this reference, the inclusion of these uncertainties reduces the expected mass-exclusion reach by approximately $0.5$~GeV, corresponding to a relative decrease of about $0.033\%$. This negligible change indicates that the expected sensitivity is predominantly statistics-limited and that the mass-exclusion reach is largely insensitive to substantial background normalization uncertainties within the adopted statistical model. This robustness test is separate from the nominal statistical analysis, which uses a $20\%$ background normalization uncertainty and a $2\%$ luminosity uncertainty.

Among the nine individual channels, the 2SSL-1J channel provides the highest sensitivity throughout the mass range considered, yielding an expected $95\%$ CL mass-exclusion reach of approximately $1375$~GeV, with $5\sigma$ ($3\sigma$) discovery reaches of $1230$ ($1410$)~GeV. It is followed by the 2OSL-$e\mu$J, 3L-1J and 3L-0J channels. The 1L channels are less competitive, primarily because of their substantially larger SM backgrounds. Nevertheless, the combined 1L channels remain more sensitive than the 4L channel up to approximately $1500$~GeV, and their sensitivity could be further improved through advanced jet-tagging techniques. 

At an integrated luminosity of $140~\mathrm{fb}^{-1}$ and $\sqrt{s}=13.6$~TeV, the present analysis yields an expected $95\%$ CL exclusion reach of approximately $1405$~GeV, with expected $5\sigma$ and $3\sigma$ discovery reaches of $1206$~GeV and $1361$~GeV, respectively. The most stringent published experimental result at comparable luminosity is the CMS search using $138~\mathrm{fb}^{-1}$ at $\sqrt{s}=13$~TeV, which reports an expected exclusion of $1060$~GeV in the FDM scenario~\cite{cms2022}. The expected exclusion reach of the present analysis exceeds this value by $345$~GeV, corresponding to a factor of $1.33$. The CMS analysis requires at least three charged leptons, whereas the present analysis additionally exploits final states with one or two leptons by reconstructing the hadronic decays of the $W$, $Z$, and $H$ bosons as Ak10 jets. The channels with fewer than three leptons account for approximately $94\%$ of the selected signal yield at $m_N=1.0$~TeV (Table~\ref{channel_yields}).

The dedicated phenomenological study of Ref.~\cite{ashanujjaman2022} provides a useful comparison, as it also considers multilepton and Ak10-jet final states at a similar collision energy of $\sqrt{s}=13$~TeV. At $500~\mathrm{fb}^{-1}$, that study reports $5\sigma$ ($3\sigma$) discovery reaches of $1265$ ($1380$)~GeV, compared with $1410$ ($1570$)~GeV obtained here, corresponding to improvements of $145$ ($190$)~GeV, or approximately $11\%$ ($14\%$). At $3000~\mathrm{fb}^{-1}$, the corresponding reaches are $1480$ ($1600$)~GeV in Ref.~\cite{ashanujjaman2022} and $1595$ ($1750$)~GeV here, corresponding to improvements of $115$ ($150$)~GeV, or approximately $8\%$ ($9\%$). These comparisons are made at identical integrated luminosities and similar collision energies. Notably, the present analysis achieves these improvements with less restrictive preselection requirements and without additional channel-specific hard cuts, relying instead on BDT-based discrimination to exploit the kinematic information and correlations among reconstructed objects and global event observables.

\section{Conclusion}\label{conclusion}

A search for heavy fermionic $\mathrm{SU}(2)_L$ triplets of the type~III seesaw mechanism has been performed using simulated $pp$ collision data at $\sqrt{s}=13.6$~TeV, corresponding to an integrated luminosity of $500~\text{fb}^{-1}$, in final states containing multiple Ak10 jets and charged leptons. The proposed search strategy, which combines orthogonal channels based on light-lepton and Ak10 jet multiplicities, significantly improves upon the sensitivity of existing multilepton-only analyses at high heavy-lepton masses. Under the flavor-democratic mixing scenario ($\mathcal{B}_e=\mathcal{B}_\mu=\mathcal{B}_\tau=1/3$), the expected $95\%$ confidence level upper limit on the triplet mass reaches approximately $1.61$~TeV, with $5\sigma$ ($3\sigma$) expected discovery reaches of $1.41$ ($1.57$)~TeV. The exclusion reach extends to approximately $1.85$~TeV at the HL-LHC luminosity of $3000$~fb$^{-1}$. These results demonstrate the strong potential of search channels combining charged leptons with boosted, hadronically decaying $W$, $Z$, and Higgs bosons to extend the discovery reach for type~III seesaw fermions at the $\sqrt{s}=13.6$~TeV LHC at an integrated-luminosity benchmark representative of the combined LHC Run~2 and Run~3 datasets, and demonstrate the potential of the proposed analysis to improve the sensitivity to type-III seesaw heavy leptons
beyond the existing experimental constraints and previous
phenomenological projections considered in this study.

\section{Data Availability}
The data that support the findings of this study are available from the corresponding author upon reasonable request.

\bibliographystyle{apsrev4-2}
\bibliography{ref}

\end{document}